\documentclass[12pt]{iopartmod}
\usepackage{graphicx}
\usepackage{iopams}

\usepackage[square,sort&compress]{natbib}
\usepackage{hyperref}

\newcommand{\D}{\ensuremath{\mathcal{D}}}
\def\vev#1{\langle #1\rangle_0}

\newcommand{\half}{\case{1}{2}}
\newcommand{\ie}{{\em i.e.}}
\newcommand{\cf}{{\em cf.\ }}

\newcommand{\gev}{\hbox{ GeV}}

\newcommand{\ev}{\hbox{ eV}}
\newcommand{\mev}{\hbox{ MeV}}

\newcommand{\tev}{\hbox{ TeV}}

\newcommand{\cm}{\hbox{ cm}}

\newcommand{\fb}{\hbox{ fb}}

\newcommand{\m}{\hbox{ m}}

\newcommand{\eqn}[1]{(\ref{#1})}
\newcommand{\Eqn}[1]{\eref{#1}}
\newcommand{\abs}[1]{\left| #1\right|}

\newcommand{\cfrac}[2]{\textstyle \frac{#1}{#2}}

\def\ltap{\mathop{\raisebox{-.4ex}{\rlap{$\sim$}} 
\raisebox{.4ex}{$<$}}}
\def\gtap{\mathop{\raisebox{-.4ex}{\rlap{$\sim$}} 
\raisebox{.4ex}{$>$}}}
\newcommand{\onetev}{1-TeV scale}
\newcommand{\lag}{\ensuremath{\mathcal{L}}}

\newcommand{\smgg}{\ensuremath{\mathrm{SU(3)}_{\mathrm{c}}\otimes\mathrm{ SU(2)_{L}} \otimes 
\mathrm{U(1)}_{Y}}}
\newcommand{\ewgg}{\ensuremath{\mathrm{SU(2)_L }\otimes \mathrm{U(1)}_Y}}

\begin{document}
\title[Origins of Mass]{Spontaneous Symmetry Breaking \\ as a Basis of Particle Mass}
\author{Chris Quigg}
\address{Theoretical Physics Department, Fermi National Accelerator Laboratory \\
P.O. Box 500, Batavia, Illinois 60510 USA
\\ and \\ Theory Group, Physics Department, CERN, CH-1211 Geneva 23, Switzerland}
\ead{\mailto{quigg@fnal.gov}}

\begin{abstract}
Electroweak theory joins electromagnetism with the weak force in a single quantum field theory, ascribing the two fundamental interactions---so different in their manifestations---to a common symmetry principle. How the electroweak gauge symmetry is hidden is one of the most urgent and challenging questions facing particle physics. The provisional answer incorporated in the ``standard model'' of particle physics was formulated in the 1960s by Higgs, by Brout \& Englert, and by Guralnik, Hagen, \& Kibble: The agent of electroweak symmetry breaking is an elementary scalar field whose self-interactions select a vacuum state in which the full electroweak symmetry is hidden, leaving a residual phase symmetry of electromagnetism. By analogy with the Meissner effect of the superconducting phase transition, the Higgs mechanism, as it is commonly known, confers masses on the weak force carriers $W^\pm$ and $Z$. It also opens the door to masses for the quarks and leptons, and shapes the world around us. It is a good story---though an incomplete story---and we do not know how much of the story is true. Experiments that explore the Fermi scale (the energy regime around $1\tev$) during the next decade will put the electroweak theory to decisive test, and may uncover new elements needed to construct a more satisfying completion of the electroweak theory. The aim of this article is to set the stage by reporting what we know and what we need to know, and to set some ``Big Questions'' that will guide our explorations.
 \end{abstract}
\pacs{12.15.-y, 14.80.Bn, 11.15.Ex\hfill \textsf{FERMILAB--PUB--07/030--T}}
\submitto{\RPP}

\maketitle

\section{Introduction to Mass \label{sec:massintro}}
In the opening lines of his \textit{Principia,}~\cite{Newton,Chandra}, Newton defines mass as ``the quantity of matter \ldots\  arising from its density and bulk conjointly.'' That intuitive notion of mass as an intrinsic attribute of matter, sharpened by $\bi{F} = m\bi{a}$ and the law of universal gravitation, is a foundation of classical physics. Mass, for Newton, is at once a measure of inertia and a source of gravitational attraction. It follows directly that mass is conserved: the mass of an object is the sum of the masses of its parts, in agreement with everyday experience. 
The extension of the law of conservation of mass to the realm of chemical reactions by Lavoisier and Lomonosov was central to the development of chemistry as a quantitative science, leading---through the work of Dalton and others---to the empirical underpinnings of the modern atomic theory. But in the classical worldview, mass does not arise, it simply is.

Mass remained an essence---part of the nature of things---for more than two centuries, until Abraham (1903) and Lorentz (1904) sought to interpret the electron mass as electromagnetic self-energy~\footnote{For Lorentz's r\'{e}sum\'{e} of  his thinking, see~\cite{Lorentz}; a modern perspective, with extensive citations to other work, appears in~\cite{ClassED}.}. Our modern conception of mass has its roots in Einstein's pregnant question~\cite{Emc2}, ``Does the inertia of a body depend upon its energy content?''  and his powerful conclusion, ``The mass of a body is a measure of its energy content; if the energy changes by $L$, the mass changes in the same sense by [$L/c^2$, where $c$ is the speed of light].'' \textit{Mass is rest-energy~\footnote{For a vivid presentation of the context and impact of the 1905 papers, see~\cite{Rigden}, and for a brisk tour of the concept of mass, see~\cite{Okun:2006pp}. To hear $E = mc^2$ pronounced in Einstein's own words, consult~\cite{MastersVoice}.}.} Among the virtues of identifying mass as $m = E_0/c^2$, where $E_0$ designates the body's rest energy, is that mass, so understood, is a Lorentz-invariant quantity, given in any frame as $m = (1/c^2)\sqrt{E^2 - p^2c^2}$. But not only is Einstein's a precise definition of mass, it invites us to consider the origins of mass by coming to terms with a body's rest energy.

For the first few steps down the quantum ladder, the difference between the Einsteinian conception of mass as rest energy and the Newtonian expectation that the mass of an object is the sum of the masses of its parts is subtle but telling. We understand the mass of an atom or molecule in terms of the masses of the atomic nuclei, the mass of the electron, and small corrections for binding energy that are given by quantum electrodynamics. ``Small corrections'' is perhaps an understatement. The 13.6-eV binding energy of the 1S electron in the hydrogen atom is but $1.45 \times 10^{-8}$ of the atom's mass. And the $13\hbox{ MJ}$ liberated in burning a cubic meter of hydrogen in the reaction 
$2\mathrm{H}_2 + \mathrm{O}_2 \to 2\mathrm{H}_2\mathrm{O}$ corresponds to a fractional mass difference between reactants and products of only $9 \times 10^{-11}$. The law of conservation of mass holds to an impressive degree;  our fossil-fuel economy feeds on the tiny deviations.

In precise and practical---if not quite ``first-principle''---terms, the masses  of all the nuclei follow from the proton mass, the neutron mass, and our semi-empirical knowledge of nuclear forces. The deeply bound $\alpha$ particle ($^4\mathrm{He}$) has a mass defect of only $\case{3}{4}\%$, so even in the nuclear realm the notion that the mass of an object is the sum of the masses of its parts is an excellent first approximation. On a macroscopic scale, that small mass defect is a proxy for a prodigious store of energy.

Nucleon mass is an entirely different story, the very exemplar of $m = E_0/c^2$. Quantum Chromodynamics, the gauge theory of the strong interactions, teaches that the dominant contribution to the nucleon mass is not the masses of the quarks that make up the nucleon, but the energy stored up in confining the quarks in a tiny volume~\cite{Wilczek:1999be}. The masses $m_u$ and $m_d$ of the up and down quarks are only a few MeV each~\cite{Yao:2006px}. The quarks contribute no more than 2\% to the 939-MeV mass of an isoscalar nucleon (averaging proton and neutron properties), because
\begin{equation}
3\,\frac{m_u + m_d}{2} = (7.5\hbox{ to }16.5)\mev.
\label{eq:Nmassfromquarks}
\end{equation}
Hadrons such as the proton and neutron thus represent \textit{matter of a novel kind.} In contrast to macroscopic matter, and to a degree far beyond what we observe in atoms, molecules, and nuclei, the mass of a nucleon is not equal to the sum of its constituent masses (up to small corrections for binding energy); \textit{it is confinement energy} (up to small corrections for constituent masses)!

QCD formulated on a spacetime lattice brings a quantitative dimension to these statements.
The CP-PACS Collaboration (centered in Tsukuba, Japan) has made a calculation omitting virtual quark-antiquark pairs that matches the observed light-hadron spectrum  at the 10\% level~\cite{Aoki:2002fd}. That discrepancy is larger than the statistical and systematic uncertainties, and so is interpreted as an artifact of the quenched (no dynamical fermions) approximation. New calculations that include virtual quark-antiquark pairs should show the full quantitative power of lattice QCD, and give us new insights into the successes and shortcomings of the simple quark model~\cite{Namekawa:2004bi,Davies:2005as}. 

The tiny $u$ and $d$ quark masses do account for an important detail of the nucleon spectrum that is essential to the world we know. The counterintuitive observation that the neutral neutron $(udd)$ is $1.29\mev$ more massive than the charged proton $(uud)$  is explained by the fact that $m_d$ exceeds $m_u$ by enough to overcome the proton's greater electromagnetic self-energy. Together with nuclear binding forces, the neutron-proton mass difference determines the pattern of radioactive decays and the roster of naturally occurring stable nuclei.

Let us be clear about the meaning of the successful calculation of the hadron spectrum using the methods of lattice QCD.  In identifying the energy of quark confinement as the origin of the nucleon mass, \textit{quantum chromodynamics has explained nearly all the visible mass of the Universe,} since the luminous matter is essentially made of protons and neutrons in stars and clouds. The oft-repeated assertion that the Higgs boson is the source of all mass in the Universe is simply incorrect---even if we restrict our attention to the luminous stuff made of ordinary baryonic matter.

The Higgs boson and the mechanism that distinguishes electromagnetism from the weak interactions are nevertheless of capital importance in shaping our world, accounting for the masses of the weak-interaction force particles and---at least in the standard electroweak theory---giving masses to the quarks and leptons. Understanding the Higgs mechanism---or whatever stands in its stead---will give us new insight into why atoms exist, how atoms can form chemical bonds, and what makes possible stable structures. These are some of the deepest questions humans have ever pursued, and they are coming within the reach of particle physics. Over the next decade, experiments will carry out definitive explorations of the Fermi scale, at energies around $1\tev$ for collisions among quarks and leptons. This \textit{nanonanoscale physics} probes distances smaller than $10^{-18}\m$, where we confidently expect to find the key to the mechanism that drives electroweak symmetry breaking. A pivotal step will be the search for the Higgs boson and the elaboration of its properties. In the same fresh terrain, we suspect that other new phenomena will give new insight into why the electroweak scale is so much smaller than the Planck scale. A class of weakly interacting dark-matter candidates could also populate the Fermi scale.

Resolving the conundrums of the Fermi scale should permit us to see more clearly the little-known territory at still shorter distances, where we may uncover new challenges to our understanding. We could well find new clues to the unification of forces or indications for a rational pattern of constituent masses, viewed at a high energy scale. If the agent of electroweak symmetry breaking turns out to be an elementary scalar, as the standard model would have it, it would be the first such particle known to experiment. Learning how it behaves could contribute important new intuition about the early-time dynamics of the inflationary Universe~\cite{GuthBook,Peacock} and the origins of dark energy~\cite{Peebles:2002gy,Feng:2005nz}.

We shall begin our tour by reviewing the standard electroweak theory, paying particular attention to the mechanism that hides the electroweak symmetry and generates masses for the weak gauge bosons, the quarks, and the leptons. The electroweak theory points to the energy scale around $1\tev$, or $10^{12}\ev$, for crucial information.  The Large Hadron Collider~\cite{lhc} soon to operate at CERN has been designed to empower experiments to carry out a thorough exploration of the Fermi scale. In the future, we expect the International Linear Collider to enrich our portrait of electroweak symmetry breaking and new phenomena on the 1-TeV scale~\cite{Dawson:2004xz}. Then we shall look briefly at a simple alternative mechanism in which the electroweak symmetry is dynamically broken by QCD. Those considerations will prepare us to uncover the broad significance of electroweak symmetry breaking by asking what the world would be like if there were no Higgs mechanism to hide the electroweak symmetry.

Next, we will consider the problem of identity---what distinguishes this quark from that lepton---and be led to take a closer look at fermion mass and mixings. We will describe signatures that will be important in the search for the Higgs boson. Then we will argue, independent of any specific mechanism for electroweak symmetry breaking, that (something like) the Higgs boson must exist. We shall find that additional considerations also single out the \onetev\ as fertile terrain for new physics.
We then look again at the idea of dynamical symmetry breaking, transplanting the QCD strategy to the Fermi scale. We shall describe a great gap in our understanding: the vacuum energy problem, and its connection to the electroweak theory. We close with a catalogue of big questions for the decade of discovery ahead.

 \section{Sources of Mass in the Electroweak Theory \label{sec:EWtheory}}
\begin{figure}[t!]
\begin{center}
\includegraphics[width=8.0cm]{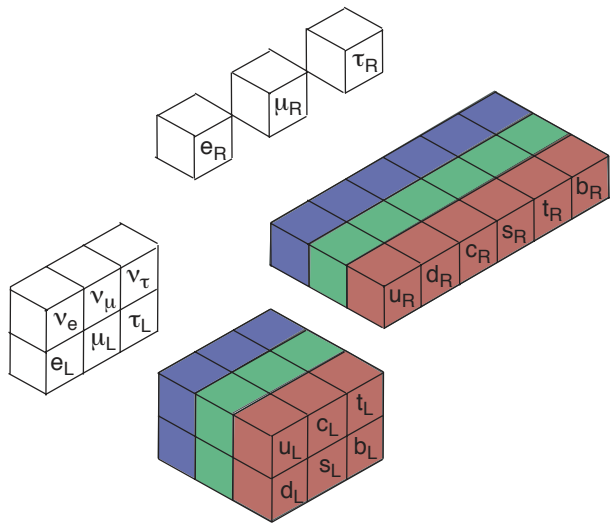}
\caption{Left-handed doublets and right-handed singlets of quarks 
and leptons that inspire 
the structure of the electroweak theory. \label{fig:DumbL}}
\end{center}
\end{figure}

 We build the standard model of particle physics~\footnote{For general surveys of the standard model of particle physics, and a glimpse beyond, see \cite{Gaillard:1998ui,CQNewPhys}.} on a set of constituents that we regard provisionally as elementary: the quarks and leptons,
as depicted in \fref{fig:DumbL}, 
plus a few fundamental forces derived from gauge symmetries. The quarks 
are influenced by the strong interaction, and so carry \textit{color}, 
the strong-interaction charge, whereas the leptons do not feel the 
strong interaction, and are colorless. We idealize the quarks and leptons as pointlike, because they  show no evidence of internal structure at 
the current limit of our resolution,  ($r \ltap 10^{-18}\m$). The charged-current weak interaction responsible for radioactive beta decay and other processes acts only on the left-handed fermions. Whether the observed parity violation reflects a fundamental asymmetry in the laws of Nature, or a left-right symmetry that is hidden by circumstance and might be restored at higher energies, we do not know. 

The electroweak theory~\footnote{Many textbooks develop the electroweak theory; see in particular~\cite{Aitchison,CQFIP56,ChengLi,Pesky}. For a look back at how the
 electroweak theory came to be, see the Nobel Lectures by some of its principal
 architects~\cite{Weinberg:1979pi,Salam:1980jd,Glashow:1979pj,'tHooft:2000xn,Veltman:2000xp}.}
 (like QCD) is a gauge theory, in which interactions follow from symmetries. Already in the 1930s,  Fermi~\cite{Enrico} and Klein~\cite{Oskar} proposed descriptions of the weak interaction in analogy to the emerging theory of quantum electrodynamics (QED).
The correct electroweak gauge symmetry, which melds the $\mathrm{SU(2)_{L}}$ family (weak-isospin) symmetry suggested by the left-handed doublets of \fref{fig:DumbL} with a $\mathrm{U(1)}_{Y}$ weak-hypercharge phase symmetry, emerged through trial and error, guided by experiment. We characterize the \ewgg\ theory by  the left-handed quarks
\begin{equation}
\mathsf{L}_q^{(1)} = 
\left(
		\begin{array}{c}
			u  \\
			d
		\end{array}
		 \right)_{\mathrm{L}} \quad
\mathsf{L}_q^{(2)} = 
		\left(
		\begin{array}{c}
			c  \\
			s
		\end{array}
		 \right)_{\mathrm{L}} \quad
		 \mathsf{L}_q^{(3)} = 
		\left(
		\begin{array}{c}
			t  \\
			b
		\end{array}
		 \right)_{\mathrm{L}}	\;,
		 \label{eq:lquarks}
	\end{equation}	
with weak isospin $I = \case{1}{2}$ and weak hypercharge $Y(\mathsf{L}_q) = \case{1}{3}$; their right-handed weak-isoscalar counterparts
\begin{equation}
\mathsf{R}_u^{(1,2,3)} = u_{\mathrm{R}}, c_{\mathrm{R}}, t_{\mathrm{R}}\hbox{ and }
\mathsf{R}_d^{(1,2,3)} = d_{\mathrm{R}}, s_{\mathrm{R}}, b_{\mathrm{R}}\;,
\label{eq:rightup}
\end{equation}
with weak hypercharges $Y(\mathsf{R}_u) = \case{4}{3}$ and $Y(\mathsf{R}_d) = -\case{2}{3}$;
the left-handed leptons
\begin{equation}
\mathsf{L}_e = 
\left(
		\begin{array}{c}
			\nu_{e}  \\
			e^{-}
		\end{array}
		 \right)_{\mathrm{L}} \;\;\;\;\;\;
\mathsf{L}_{\mu} = 
		\left(
		\begin{array}{c}
			\nu_{\mu}  \\
			\mu^{-}
		\end{array}
		 \right)_{\mathrm{L}} \;\;\;\;\;\;
\mathsf{L}_{\tau} = 
		\left(
		\begin{array}{c}
			\nu_{\tau}  \\
			\tau^{-}
		\end{array}
		\right)_{\mathrm{L}}	\;,
		\label{eq:lleptons}
	\end{equation}
with weak isospin $I = \case{1}{2}$ and weak hypercharge $Y(\mathsf{L}_{\ell}) = -1$; and the right-handed weak-isoscalar charged leptons
\begin{equation}
\mathsf{R}_{e,\mu,\tau} = e_{\mathrm{R}}, \mu_{\mathrm{R}}, \tau_{\mathrm{R}}\;,
\label{eq:rightlep}
\end{equation}
with weak hypercharge $Y(\mathsf{R}_{\ell}) = -2$. (Weak isospin and weak hypercharge are related to electric charge through $Q = I_{3} + \cfrac{1}{2}Y$.) Here we have idealized the neutrinos as massless; we will touch on possible sources of neutrino mass in \sref{sec:fermions2}.

The \ewgg\ electroweak gauge group implies two sets of gauge fields: a weak isovector $\bi{b}_\mu$, with coupling constant $g$, and a weak isoscalar ${{\mathcal A}}_\mu$, with independent coupling constant $g^\prime$. The gauge fields compensate for the variations induced by gauge transformations, provided that they obey the transformation laws $\bi{b}_\mu \to \bi{b}_\mu - \balpha \times \bi{b}_\mu - (1/g)\partial_\mu \balpha$ under an infinitesimal weak-isospin rotation generated by $G = 1 + (\rmi/2)\balpha \cdot \btau$ (where $\btau$ are the Pauli isospin matrices) and $\mathcal{A}_\mu \to \mathcal{A}_\mu - (1/g^\prime)\partial_\mu \alpha$ under an infinitesimal hypercharge phase rotation.
Corresponding to these gauge fields are the field-strength tensors 
\begin{equation}
    F^{\ell}_{\mu\nu} = \partial_{\nu}b^{\ell}_{\mu} - 
    \partial_{\mu}b^{\ell}_{\nu} + 
    g\varepsilon_{jk\ell}b^{j}_{\mu}b^{k}_{\nu}\; ,
    \label{eq:Fmunu}
\end{equation}
for the weak-isospin symmetry, and 
\begin{equation}
    f_{\mu\nu} = \partial_{\nu}{{\mathcal A}}_\mu - \partial_{\mu}{{\mathcal 
    A}}_\nu \; , 
    \label{eq:fmunu}
\end{equation}
for the weak-hypercharge symmetry. 
 
We may summarize the interactions 
by the Lagrangian
\begin{equation}
\lag = \lag_{\rm gauge} + \lag_{\rm leptons} +  \lag_{\rm quarks}\ ,                           
\end{equation}             
with
\begin{equation}
\lag_{\rm gauge}=-\cfrac{1}{4}\bi{F}_{\mu\nu} \cdot  \bi{F}^{\mu\nu}
-\cfrac{1}{4}f_{\mu\nu}f^{\mu\nu},
\label{eq:gaugeL}
\end{equation}
\begin{eqnarray}     
\lag_{\rm leptons} & = & \overline{{\sf R}}_{\ell}\:\rmi\gamma^\mu\!\left(\partial_\mu
+\rmi\frac{g^\prime}{2}{\cal A}_\mu Y\right)\!{\sf R}_{\ell}
\label{eq:matiere} \\ 
& + & \overline{{\sf
L}}_{\ell}\:\rmi\gamma^\mu\!\left(\partial_\mu 
+\rmi\frac{g^\prime}{2}{\cal
A}_\mu Y+\rmi\frac{g}{2}\btau\cdot\bi{b}_\mu\right)\!{\sf L}_{\ell}\;, \nonumber
\end{eqnarray}
where $\ell$ runs over $e, \mu, \tau$, and
\begin{eqnarray}     
\lag_{\rm quarks} & = & \overline{{\sf R}}_{u}^{(n)}\:\rmi\gamma^\mu\!\left(\partial_\mu
+\rmi\frac{g^\prime}{2}{\cal A}_\mu Y\right)\!{\sf R}_{u}^{(n)}
\nonumber \\ 
 & + & \overline{{\sf R}}_{d}^{(n)}\:\rmi\gamma^\mu\!\left(\partial_\mu
+\rmi\frac{g^\prime}{2}{\cal A}_\mu Y\right)\!{\sf R}_{d}^{(n)}
\label{eq:qmatiere}  \\
& + & \overline{{\sf
L}}_{q}^{(n)}\:\rmi\gamma^\mu\!\left(\partial_\mu 
+\rmi\frac{g^\prime}{2}{\cal
A}_\mu Y+\rmi\frac{g}{2}\btau\cdot\bi{b}_\mu\right)\!{\sf L}_{q}^{(n)}\;, \nonumber
\end{eqnarray}
where $n$ runs over $1, 2, 3$.

Although the weak and electromagnetic interactions share a common origin in the \ewgg\ gauge symmetry, their manifestations are very different. Electromagnetism is a force of infinite range, while the influence of the charged-current weak interaction responsible for radioactive beta decay only spans distances shorter than about $10^{-15}\cm$. The phenomenology is thus at odds with the theory we have developed to this point. The gauge Lagrangian \eref{eq:gaugeL} contains four massless electroweak gauge bosons, namely ${{\mathcal A}}_\mu$, $b^{1}_{\mu}$, $b^{2}_{\mu}$, and $b^{3}_{\mu}$, because a mass term such as $\cfrac{1}{2}m^2\mathcal{A}_\mu\mathcal{A}^\mu$ is not invariant under a gauge transformation. Nature has but one: the photon. Moreover, the \ewgg\ gauge symmetry forbids fermion mass terms $m\bar{f}\!f = m(\bar{f}_{\mathrm{R}}f_{\mathrm{L}} + \bar{f}_{\mathrm{L}}f_{\mathrm{R}})$ in \eref{eq:matiere} and \eref{eq:qmatiere}, because the left-handed and right-handed fields transform differently. 

To give masses to the gauge bosons and constituent fermions, we must hide the electroweak symmetry, recognizing that a symmetry of the laws of Nature does not imply that the same symmetry will be manifest in the outcomes of those laws. How the electroweak gauge symmetry is spontaneously broken---hidden---to the $\mathrm{U(1)_{\mathrm{em}}}$ phase symmetry of electromagnetism is one of the most urgent and challenging questions before particle physics.
   
The superconducting phase transition offers an instructive model for hiding the electroweak gauge symmetry\footnote{See  \S 4.4 of~\cite{Marshak} and \S 21.6 of~\cite{SWFT}.}. To give masses to the intermediate bosons of the weak interaction, we appeal to the Meissner effect---the exclusion of magnetic fields from a superconductor, which corresponds to the photon developing a nonzero mass within the superconducting medium. What has come to be called the Higgs mechanism~\cite{Higgs:1964ia,Englert:1964et,Higgs:1964pj,Guralnik:1964eu,Higgs:1966ev,PhysRev.155.1554} is a relativistic generalization of the Ginzburg-Landau phenomenology~\cite{Ginzburg:1950sr} of superconductivity\footnote{Early steps toward understanding spontaneously broken gauge symmetry are recalled in~\cite{TWBK,GHKrev,Higgs:2002ht,Brout:1998qb,Lee:1977ur,Aitchison:1989mw,Weinberg:2004kv}.}. The essential insight is that the Goldstone theorem~\cite{Goldstone:1961eq,Goldstone:1962es}~\footnote{If the Lagrangian of a local, manifestly Lorentz-invariant quantum field theory with positive-definite metric on the Hilbert space is invariant under a continuous symmetry, then either the vacuum state is also invariant under that symmetry or a massless spin-zero ``Goldstone boson'' corresponds to each broken generator. Nonrelativistic evasions are exhibited in~\cite{PhysRev.117.648,PhysRev.130.439,PhysRevLett.12.266}.} does not operate when a local gauge symmetry, as opposed to a continuous global symmetry, is broken. Instead, a miraculous interplay between the would-be Goldstone bosons and the normally massless gauge bosons endows gauge bosons with mass and removes the massless scalars from the spectrum. 

Let us see how spontaneous symmetry breaking operates in the electroweak theory.
We introduce a complex doublet of scalar fields
\begin{equation}
\phi\equiv \left(\begin{array}{c} \phi^+ \\ \phi^0 \end{array}\right)
\end{equation}
with weak hypercharge $Y_\phi=+1$.  Next, we add to the Lagrangian new 
(gauge-invariant) terms for the interaction and propagation of the 
scalars,
\begin{equation}
      \lag_{\rm scalar} = (\D^\mu\phi)^\dagger(\D_\mu\phi) - V(\phi^\dagger \phi),
\end{equation}
where the gauge-covariant derivative is
\begin{equation}
      \D_\mu=\partial_\mu 
+\rmi\frac{g^\prime}{2}{\cal A}_\mu
Y+\rmi\frac{g}{2}\btau\cdot\bi{b}_\mu \; ,
\label{eq:GcD}
\end{equation}
and (inspired by Ginzburg \& Landau) the potential interaction has the form
\begin{equation}
      V(\phi^\dagger \phi) = \mu^2(\phi^\dagger \phi) +
\abs{\lambda}(\phi^\dagger \phi)^2 .
\label{SSBpot}
\end{equation}
We are also free to add  gauge-invariant Yukawa interactions between the scalar fields
and the leptons ($\ell$ runs over $e, \mu, \tau$ as before),
\begin{equation}
      \lag_{{\rm Yukawa-}\ell} = -\zeta_{\ell}\left[(\overline{{\sf L}}_{\ell}\phi){\sf R}_{\ell} + 
      \overline{{\sf R}}_{\ell}(\phi^\dagger{\sf
L}_{\ell})\right]\;,
\label{eq:Yukterm}
\end{equation}
and similar interactions with the quarks---about which we shall have more to say presently, in \sref{sec:fermions2}.

We then arrange 
their self-interactions so that the vacuum state corresponds to a 
broken-symmetry solution.  The electroweak symmetry is spontaneously broken if the parameter
$\mu^2$ is taken to be negative. In that event, gauge invariance gives us the freedom to choose the state of minimum energy---the vacuum state---to correspond to the vacuum expectation value
\begin{equation}
\vev{\phi} = \left(\begin{array}{c} 0 \\ v/\sqrt{2} \end{array}
\right),
\label{eq:vevis}
\end{equation}
where $v = \sqrt{-\mu^2/\abs{\lambda}}$.

Let us verify that the vacuum of \eref{eq:vevis} does break the gauge symmetry $\ewgg \to \mathrm{U(1)}_{\mathrm{em}}$.  The vacuum state $\vev{\phi}$ is invariant under a symmetry operation corresponding to the generator ${\mathcal G}$ provided that $\rme^{\rmi \alpha {\mathcal G}}\vev{\phi} = \vev{\phi}$, \ie, if ${\mathcal G}\vev{\phi} = 0$.  
Direct calculation reveals that  the  original four generators are all broken, but electric charge is
not.  The photon remains massless, but the other three gauge bosons acquire 
masses, as auxiliary scalars assume the role of the third 
(longitudinal) degrees of freedom.  

Introducing the weak mixing angle $\theta_{W}$ and defining $g^{\prime} = g\tan\theta_{W}$, we can express the photon as the linear combination
$A = \mathcal{A}\cos{\theta_{W}} + b_{3}\sin{\theta_{W}}$. We identify the strength of its (pure vector) coupling to charged particles, $gg^{\prime}/\sqrt{g^{2} + g^{\prime 2}}$, with the electric charge $e$.
The mediator of the charged-current weak 
interaction, $W^{\pm} = (b_{1} \mp ib_{2})/\sqrt{2}$, acquires a 
mass $M_{W} = gv/2 = ev/2\sin{\theta_{W}}$. The electroweak gauge theory reproduces the low-energy phenomenology of the $V - A$ theory of weak interactions, provided we set $v = (G_{\mathrm{F}}\sqrt{2})^{-1/2} = 246\gev$, where $G_{\mathrm{F}} = 1.166 37(1) \times 10^{-5}\gev^{-2}$ is Fermi's weak-interaction coupling constant. It follows at once that $M_W \approx 37.3\gev/\sin{\theta_{W}}$. The combination of the $I_3$ and $Y$ gauge bosons orthogonal to the photon is the mediator of the neutral-current weak interaction, $Z = b_{3}\cos{\theta_{W}} - \mathcal{A}\sin{\theta_{W}}$, which
acquires a mass $M_Z=M_W/\cos{\theta_W}$. The weak neutral-current interaction was not known before the electroweak theory. Its discovery in 1973~\cite{Hasert:1973ff,Haidt:2004ne} marked an important milestone, as did the observation a decade later~\cite{Rubbia:1985pv} of the $W^{\pm}$~\cite{Arnison:1983rp,Banner:1983jy} and $Z^0$~\cite{Arnison:1983mk,Bagnaia:1983zx} bosons.

Three decades of extensive studies of the weak neutral current culminated in experiments at the $e^+e^- \to Z$ factories. The ALEPH, DELPHI, L3, and OPAL detectors accumulated 17 million $Z$ decays  at LEP, and the SLD detector recorded 600 thousand $Z$ decays using polarized beams at the Stanford Linear Collider~\cite{Z-Pole}.
A broad collection of experimental measurements and the supporting theoretical calculations have elevated the electroweak theory to a law of Nature, tested as a quantum field theory at the level of one part in a thousand~\cite{deJong:2005mk, lepewwg,Yao:2006px,unknown:2006mx}. The mass of the neutral weak boson is known to impressive precision, $M_Z =  91.1876 \pm  0.0021\gev$, while the world average $W$-boson mass is $M_W = 80.398 \pm 0.025\gev$~\cite{Run2W}. By themselves, these measurements imply an ``on-shell'' weak mixing parameter $\sin^2\theta_W \equiv 1 - M_W^2/M_Z^2 = 0.22265 \pm 0.00052$, in fine agreement with the Particle Data Group 2006 grand average, $0.22306 \pm 0.00031$~\cite{Yao:2006px}. The quantum (loop) corrections to many observables, including the ratio $M_W/M_Z$, are sensitive to the top-quark mass, and showed a preference for a heavy top before the discovery. Now the comparison of indirect inferences with the measured top-quark mass is one of many consistency checks for the electroweak theory. The 2006 inferred value, $m_t^{(\mathrm{indirect)}} = 172.3^{+10.2}_{-7.6}\gev$~\cite{Yao:2006px} matches the (more precise) Tevatron average of direct measurements, $m_t = 170.9 \pm 1.8\gev$~\cite{TevTop0307}. 

\begin{figure}[tb]
	\centerline{\includegraphics[width=8.0cm]{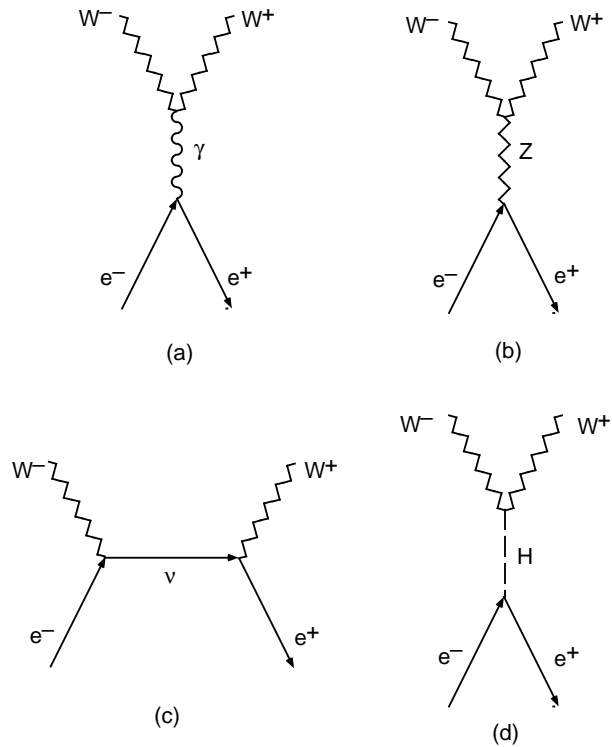}}
	\vspace*{6pt}
	\caption{Lowest-order contributions to the $e^+e^- \rightarrow 
	W^{+}W^{-}$ scattering amplitude.}
	\protect\label{fig:eeWW}
\end{figure}

\begin{figure}[tb]
	\centerline{\includegraphics[width=8.0cm]{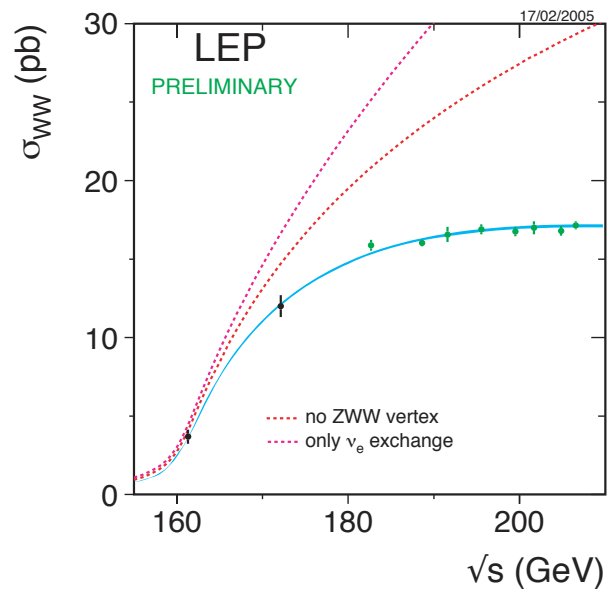}}
	\vspace*{6pt}
	\caption{Cross section for the reaction $e^{+}e^{-} \to W^{+}W^{-}$ 
	measured by the four LEP experiments, together with the full 
	electroweak-theory simulation and the cross sections that would 
	result from $\nu$-exchange alone and from $(\nu+\gamma)$-exchange
	{\protect \cite{lepewwg}}.}
	\protect\label{fig:LEPgc}
\end{figure}

 One noteworthy achievement is a clear 
test of the electroweak gauge symmetry in the reaction $e^{+}e^{-} \to W^{+}W^{-}$. 
Neglecting the electron mass, this reaction is described by three 
Feynman diagrams that correspond to  $s$-channel photon and $Z^{0}$ exchange, and $t$-channel neutrino exchange, \fref{fig:eeWW}(a-c). 
For the production of longitudinally polarized $W$-bosons, each diagram leads to a $J = 1$ partial-wave amplitude that grows as the square of the c.m.\ energy, but the gauge symmetry enforces a pattern of cooperation.
 The contributions of the direct-channel 
$\gamma$- and $Z^0$-exchange diagrams 
of \fref{fig:eeWW}(a) and (b) cancel the leading divergence in the $J=1$ 
partial-wave amplitude of the neutrino-exchange diagram in \fref{fig:eeWW}(c).  The interplay is shown in \fref{fig:LEPgc}. If the $Z$-exchange contribution is omitted (middle line) or if both the 
$\gamma$- and $Z$-exchange contributions are omitted (upper line), the calculated cross section grows unacceptably with 
energy. The measurements compiled by the LEP Electroweak Working Group~\cite{lepewwg} 
agree well with the benign high-energy behavior predicted by the full electroweak theory, confirming the gauge cancellation in the $J=1$ partial-wave amplitude.

Three of the four scalar degrees of freedom that we introduced to forge a vacuum state that does not manifest the gauge symmetry have become the longitudinal components of $W^+$, $W^-$, and $Z$. What of the fourth? It appears as a vestige 
of the spontaneous symmetry breaking, in the form of a massive
spin-zero particle, called the Higgs boson, $H$.  Its mass  is 
given symbolically as $M_{H}^{2} = -2\mu^{2} > 0$, but \textit{we have no 
prediction for its value.}  What we take to be the work of the 
Higgs boson is all around us \footnote{We shall see in \sref{sec:higgsprops} that the Higgs-boson contribution of \fref{fig:eeWW}(d) ensures acceptable high-energy behavior of the $J=0$ partial-wave amplitude.} , as we shall detail below, but  the Higgs particle itself has not yet 
been observed!

\begin{figure}[t!]
\begin{center}
\includegraphics[width=8.0cm]{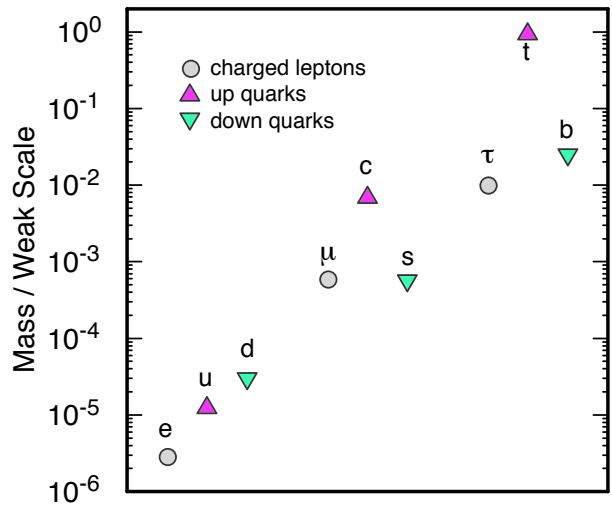}
\caption{Yukawa couplings $\zeta_i = m_i / (v/\sqrt{2})$ inferred from the fermion masses in \tref{table:masses}. \label{fig:yuk}}
\end{center}
\end{figure}

The masses of the elementary fermions collected in \tref{table:masses} 
\begin{table}[tb]
\caption{Masses of the charged leptons and quarks~\cite{Yao:2006px}.}
\begin{center}
\begin{tabular}{|c|c|}
\hline
Particle &	Mass [MeV]\\
\hline
$e$ &	$0.51099892 \pm 0.00000004$ \\
$\mu$ &	$105.658369 \pm 0.000009$ \\
$\tau$ &	$1\,776.99^{+0.29}_{-0.26}$ \\[3pt]
$u$ &	$2.25 \pm 0.75$ \\
$d$	& $5.00 \pm 2.00$ \\
$s$	& $95 \pm 25$ \\
$c$	& $1\,250 \pm 90$ \\
$b$ &	$4\,200 \pm 70$ \\
$t$	& $170\,900 \pm 1\,800$~\cite{TevTop0307} \\
\hline
\end{tabular}
\end{center}
\label{table:masses}
\end{table}%
are a more mysterious 
story. Each fermion mass involves a new Yukawa coupling $\zeta$ (\cf \eref{eq:Yukterm}).  
When the electroweak symmetry is spontaneously broken, the electron 
mass emerges as $m_{e} = \zeta_{e}v/\sqrt{2}$. The Yukawa couplings that reproduce the observed quark and lepton masses range over many orders of magnitude, as shown in \fref{fig:yuk}.
The origin of the Yukawa couplings is obscure: they do not follow from a known symmetry principle, for example. In that sense, therefore, \textit{all fermion masses involve physics
beyond the standard model.} 

Let us summarize what we have learned about the sources of particle mass in the standard electroweak theory. Unless the electroweak gauge symmetry is hidden, the four gauge bosons and all the constituent fermions are massless. Spontaneous symmetry breaking, in the form of the Higgs mechanism, gives masses to the weak gauge bosons and creates the possibility for the fermions to acquire mass. Once the weak mixing parameter $\sin^2\theta_W$ is fixed by the study of weak-neutral-current interactions, the theory makes successful quantitative predictions for the $W^\pm$- and $Z$-boson masses. Although the natural scale of fermion masses would seem to be set by the electroweak scale, the masses themselves are determined by Yukawa couplings of the fermions to the Higgs field. We do not know what fixes the size of the Yukawa couplings. Finally, the theory requires a scalar Higgs boson, but does not make an explicit prediction for its mass.

 \section{The Significance of the Fermi Scale \label{sec:import}}
The electroweak theory does not make a testable prediction for the Higgs-boson mass because we do not know the values of the Higgs-potential parameters $\abs{\lambda}$ and $\mu^2$ that enter the symbolic prediction $M_H^2 = -2\mu^2 = 2 \abs{\lambda}v^2$. However, a thought experiment~\cite{Lee:1977eg} leads to a conditional upper bound on the Higgs-boson mass that identifies a high-value experimental target. 

Imagine pairwise collisions among the $W^\pm$ bosons that mediate charge-changing weak interactions, the $Z$ boson responsible for weak neutral-current interactions, and the Higgs boson $H$.
 Most channels ``decouple,'' in the sense 
that partial-wave amplitudes are small at all energies (except very
near the particle poles, or at exponentially large energies), for
any value of the Higgs boson mass $M_H$. Four neutral channels---$W_0^+W_0^-$, $Z_0Z_0$, $HH$, and $HZ_0$, where the subscript $0$ denotes longitudinal polarization---are particularly interesting. Amplitudes calculated at lowest order in the electroweak theory make sense in the limit of high energies, in that the probability of a scattering event does not exceed unity, provided that the Higgs-boson mass is not too large. Specifically, the $J=0$ partial-wave amplitudes involving these channels  are all asymptotically
constant (\ie, well-behaved) and  
proportional to Fermi's constant $G_{\mathrm{F}}$ times the square of the Higgs-boson mass.
Requiring that the largest eigenvalue respect the 
partial-wave unitarity condition $\abs{a_0}\le 1$ yields
\begin{equation}
	M_H \le \left(\frac{8\pi\sqrt{2}}{3G_{\mathrm{F}}}\right)^{1/2} \approx 1\tev\;,
\end{equation}
which characterizes the Fermi scale.

If $M_H$ respects the bound, weak interactions remain weak at all energies, and perturbation theory is everywhere reliable. If the Higgs boson were heavier than $1\tev$, the weak interactions among $W^\pm$, $Z$, and $H$ would become strong on the Fermi scale, and perturbation theory would break down. At TeV energies, we might then observe multiple production of weak bosons, $W^+W^-$ resonances, and other phenomena evocative of pion-pion scattering at GeV energies~\cite{Chanowitz:2004gk}. One way or another, something new---a Higgs boson or strong scattering, if not some other new physics~\cite{Csaki:2005vy,Schmaltz:2005ky}---is to be found in electroweak interactions at energies not much larger than $1\tev$~\cite{Veltman:1976rt,Lee:1977eg,Dicus:1992vj}.

\section{A Dynamical Approach to Electroweak Symmetry Breaking \label{sec:techni}}
The analogy between electroweak symmetry breaking and the superconducting phase transition led to the insight of the Higgs mechanism. The macroscopic order parameter of the Ginzburg-Landau phenomenology, which corresponds to the wave function of superconducting 
charge carriers, acquires a nonzero vacuum expectation value in the 
superconducting state. The microscopic Bardeen-Cooper-Schrieffer 
theory \cite{Bardeen:1957kj} interprets the dynamical origin of the order parameter through 
the formation of correlated states of elementary fermions, the Cooper pairs of 
electrons. 

The elementary fermions---electrons---and 
gauge interactions---QED---needed to generate the correlated pairs are 
already present in the case of superconductivity. Could a scheme
 of similar economy account
for the transition that hides the electroweak symmetry?
Consider an \smgg\ theory of massless up and 
down quarks. Because the strong interaction is strong and the electroweak 
interaction is feeble we may treat the \ewgg\
interaction as a perturbation. For vanishing quark masses, QCD displays an exact 
$\mathrm{SU(2)_L\otimes SU(2)_R}$ chiral symmetry. At an energy scale 
$\sim\Lambda_{\mathrm{QCD}},$ the strong interactions become strong, fermion 
condensates appear, and the chiral symmetry is spontaneously broken
to the familiar flavor symmetry, isospin:
\begin{equation}
	\mathrm{SU(2)_L\otimes SU(2)_R \to SU(2)_V}\;\; .
\end{equation}
 Three Goldstone bosons appear, one for 
each broken generator of the original chiral invariance. These were 
identified by Nambu~\cite{Nambu:1960xd} as three massless pions.

The broken generators are three axial currents whose couplings to pions are 
measured by the pion decay constant $f_\pi$, which is measured by the charged-pion lifetime. When we turn on the 
\ewgg\ electroweak interaction, the electroweak gauge 
bosons couple to the axial currents and acquire masses of order $\sim 
gf_\pi$. The mass-squared matrix,
\begin{equation}
	\mathcal{M}^{2} = \left(
		\begin{array}{cccc}
		g^{2} & 0 & 0 & 0  \\
		0 & g^{2} & 0 & 0  \\
		0 & 0 & g^{2} & gg^{\prime}  \\
		0 & 0 & gg^{\prime} & g^{\prime2}
	\end{array}
		 \right) \frac{f_{\pi}^{2}}{4} \; ,
	\label{eq:csbm2}
\end{equation}
(where the rows and columns correspond to $W_1$, $W_2$, $W_{3}$, 
and $\mathcal{A}$) has the same structure as the mass-squared matrix 
for gauge bosons in the standard electroweak theory.  Diagonalizing 
the matrix \eref{eq:csbm2}, we find that the photon, corresponding as in the standard model to the combination $A = (g\mathcal{A} + g^{\prime}b_3)/\sqrt{g^2 + g^{\prime 2}}$, emerges massless.
Two charged gauge bosons, $W^{\pm} = (b_1 \mp \rmi b_2)/\sqrt{2}$, acquire mass-squared
$M_{W}^{2} = g^{2}f_{\pi}^{2}/4$, and the neutral gauge boson $Z = (-g^{\prime}\mathcal{A} + gb_3)/ \sqrt{g^2 + g^{\prime 2}}$ obtains $M_{Z}^{2} = 
(g^{2}+g^{\prime2})f_{\pi}^{2}/4$. The ratio,
\begin{equation}
	\frac{M_{Z}^{2}}{M_{W}^{2}} = \frac{(g^{2}+g^{\prime2})}{g^{2}} = 
	\frac{1}{\cos^{2}\theta_{W}}\; ,
	\label{eq:wzrat}
\end{equation}
reproduces the standard-model result.

The massless pions thus disappear from the physical spectrum, 
having become the longitudinal components of the weak gauge bosons. 
Despite the structural similarity to the standard model, the chiral symmetry breaking of QCD does not yield a satisfactory theory of the weak interactions. The masses acquired by the 
intermediate bosons are $2\,500$ times smaller than required for a successful 
low-energy phenomenology; the $W$-boson mass is only~\cite{Weinstein:1973gj} $M_W\approx 30\mev$, because its scale is set by $f_{\pi}$.

 \section{A World without the Higgs Mechanism \label{sec:itmatters}}
Exploring the Fermi scale will bring us a new appreciation of what lies behind the complexity and diversity of the everyday world. To see what we can hope to learn, let us consider how the world would be changed if we could not rely on something like the Higgs mechanism to hide electroweak symmetry.

What a different world it would be! 
First, the quarks and leptons would
remain massless, because mass terms are not permitted  if the electroweak symmetry remains
manifest. Eliminating the Higgs mechanism does nothing to alter the strong interaction, so QCD would
still confine the (now massless) color-triplet quarks into color-singlet
hadrons, now with many ``light hadrons'' because there are no heavy quarks.  

If the electroweak symmetry were unbroken, the asymptotically free weak-isospin force would confine objects that carry weak isospin into weak-isospin singlets. But as we have just seen in \sref{sec:techni}, the chiral condensate of QCD hides the electroweak symmetry, even in the absence of a Higgs mechanism. Because the weak bosons have acquired mass, the $\mathrm{SU(2)_L}$ interaction does not confine. The familiar light-hadron spectrum persists, but with a crucial difference. In the no-Higgs-boson scenario, with no $u$-$d$ quark mass difference to tip the balance, the proton would outweigh the neutron. The pattern of radioactive beta decay would be turned on its head. In our world, a free neutron decays ($n \to p e^- \bar{\nu}_e$) with a mean life of about fifteen minutes. If quark masses vanish and $M_W \approx 30\mev$, a free proton decays in less than a millisecond: $p \to n e^+ \nu_e$. There is no hydrogen atom, and the lightest ``nucleus'' would be one neutron.

It seems likely that some light elements would be produced in the early no-Higgs
universe~\cite{PhysRevLett.80.1822,PhysRevD.57.5480,RevModPhys.72.1149,PhysRevD.67.043517}. But even if some nuclei are produced and survive, they would not form atoms we would recognize. A massless electron means that the Bohr radius of an atom---half a nanometer in our world---would be infinite. [Now, it is nearly inevitable that effects negligible in our world would, in the Higgsless world, produce fermion masses many orders of magnitude smaller than those we observe. The Bohr radius of a would-be atom would be macroscopic, sustaining the conclusion that matter would lose its integrity.]

A world without compact atoms would be a world without chemical valence bonds and without stable composite structures like our solids and liquids. All matter would be insubstantial---and \textit{we} would not exist!

\section{The Problem of Identity \label{sec:identity}}
Contemplating the variety of the quarks and leptons invites the tantalizing 
question, ``What makes a top quark a top quark, an electron an 
electron, and a neutrino a neutrino?'' In more operational terms, we may ask, ``What determines 
the masses and mixings of the quarks and leptons?'' It is not enough 
to answer, ``The Higgs mechanism,'' because the fermion masses are 
a very enigmatic element of the electroweak theory.  Once the electroweak 
symmetry is hidden, the electroweak theory permits---welcomes---fermion masses, 
but the values of the masses are set by the 
couplings of the Higgs boson to the fermions, which are of unknown provenance. Nothing 
in the electroweak theory is ever going to prescribe those couplings. 
It is not that the calculation is technically challenging; \textit{there is 
no calculation} \footnote{Either family symmetries or unified theories of the strong, weak, and electromagnetic interactions, in which quarks and leptons are members of extended families, offer the prospect of simple relations among fermion masses at a high-energy scale. These relations, which reflect the symmetries and the pattern of symmetry breaking, are modulated by the running of the masses down to the low-energy scales on which we measure them~\cite{Chen:2003zv,Raby:2004br,Mohapatra:2006gs}. Models that incorporate extra spacetime dimensions present new ways to think about the exponential range of Yukawa couplings.  If the standard-model fields were constrained to a thick wall, the wave packets representing different fermion species might be fixed on different tracks within the extra dimension~\cite{Arkani-Hamed:1999dc,Mirabelli:1999ks}. Yukawa couplings would measure the extra-dimensional overlap of the left-handed and right-handed fermion wave packets and the Higgs field, presumed pervasive.  Small offsets in the new coordinate could yield exponentially large mass differences.}.

The exciting prospect, then, is that quark and lepton masses, mixing 
angles, and the subtle differences in the behavior of particles and their antiparticles manifested as \textsf{CP} violation put us in contact with 
physics beyond the standard model. The challenge is to 
construct what the big question really is. We may find new phenomena that suggest the origin of some or all of the quark and lepton masses. The extremely light ($\ltap 2\ev$) neutrinos---which might be their own antiparticles, because they are electrically neutral---may be special, acquiring some or all of their mass from a mechanism not open to the quarks and charged leptons. And it might just be that we haven't 
recognized a latent pattern in the masses because we're not seeing the 
whole picture yet. Perhaps it will take discovering the spectrum of a new kind of 
matter---superpartners, or something entirely different---before it all begins to make 
sense.

Should we expect to find an ultimate resolution to the problem of identity?
According to a recurring dream for generations of physicists, the theory of the world 
    might prove to be so restrictive that things have to turn out the 
    way we observe them. Is this really the way the world works, or 
    not? Are the elements of our standard model---the quarks and 
    leptons and gauge groups and coupling constants---inevitable, at 
    least in a probabilistic sense, or did it just happen this 
    way?
 
 It may be instructive to call to mind Johannes Kepler's quest to understand why the
Sun should have exactly six planetary companions in the observed
orbits~\cite{LincolnWolfenstein04292003}.  Kepler sought a symmetry principle that would give
order to the universe following the Platonic-Pythagorean tradition.
Perhaps, he thought, the six orbits were determined by the five regular solids of
geometry, or perhaps by musical harmonies.  We now know that the Sun
holds in its thrall more than six planets, not to mention the
asteroids, periodic comets, and planetini, nor all the moons around
Kepler's planets.  But that is not why Kepler's problem seems
ill-conceived to us; we just do not believe that it should have a
simple answer. Neither symmetry principles nor stability criteria make
it inevitable that those six planets should orbit our Sun precisely 
as they do. This example holds two lessons for us: First, it 
is very hard to know in advance which aspects of the physical world 
will have simple, beautiful, informative explanations, and which we 
shall have to accept as ``complicated,'' or environmental parameters. Second, and here Kepler is a 
particularly inspiring example, we may learn very great lessons 
indeed while pursuing challenging questions that---in the end---do not 
have illuminating answers.

Are Nature's Laws the same at all times and places? Yes, of course they are, \textit{to good approximation, in our experience.} Otherwise science would have had to confront a universe that is in some manner capricious. But \textit{all} times and \textit{all} places is a very strong conclusion, for which we cannot have decisive evidence. Some cosmologists argue that our Universe is but one trifling pocket within a multiverse of mind-boggling proportions. Proponents of the string-theory landscape---a mathematical meta-space representing all the alternative environments that theory allows---see a grand panorama of self-consistent possibilities in which what we take to be the laws of nature apply only in our corner. Everywhere in the landscape, the home team plays by its own rules of physics, derived from the locally prevalent gauge symmetries, elementary-particle spectra, and coupling constants~\cite{Susskind:2003kw,Lenny,WeinbergMulti}.

An eventual explanation of masses and mixings and \textsf{CP} violation might come in the form of inevitability, or probability, or possibility. Exploring the Fermi scale, it seems to me, is highly likely to resolve the question of mechanism often attributed to Richard Feynman,  ``\textit{Why} does the muon weigh?'' The follow-up question,  ``\textit{What} does the muon weigh?'' may be trickier to settle.

\section{A closer look at fermion masses \label{sec:fermions2}}
In the standard electroweak theory, the quarks and leptons are taken to be elementary particles. The masses and mixings of the quarks arise from Yukawa interactions with the Higgs condensate,
\begin{equation}
\mathcal{L}_{\mathrm{Yukawa}} = -\zeta_d^{ij}(\bar{\mathsf{L}}_i \phi)d_{\mathrm{R}j} 
-\zeta_u^{ij}(\bar{\mathsf{L}}_i \bar{\phi})u_{\mathrm{R}j} + \hbox{h.c.}\;,
\label{eqn:quarkyuk}
\end{equation}
where the Yukawa couplings $\zeta_{u,d}$ are $3\times 3$ complex matrices, $i$ and $j$ are generation indices, $\mathsf{L}_i$ are left-handed quark doublets, $u_{\mathrm{R}j}$ and $d_{\mathrm{R}j}$ are right-handed quark singlets, and $\bar{\phi} = \rmi \sigma_2 \phi^*$. \Eref{eqn:quarkyuk} yields quark mass terms when the Higgs field $\phi$ acquires a vacuum expectation value \eref{eq:vevis}, as we saw in the penultimate paragraph of \sref{sec:EWtheory}. The mass eigenstates are obtained by diagonalizing the Yukawa matrices $\zeta^{\mathrm{diag}}_f = U^f_{\mathrm{L}}\zeta_f U^{f\dagger}_{\mathrm{R}}$, where $f = u,d$ refers to up-like or down-like quarks and $U^f_{\mathrm{L,R}}$ are unitary matrices (\cf\ \fref{fig:yuk}). Accordingly, the charged-current interactions among the left-handed quarks $\bi{u}_{\mathrm{L}} = (u_{\mathrm{L}}, c_{\mathrm{L}}, t_{\mathrm{L}})$ and $\bi{d}_{\mathrm{L}} = (d_{\mathrm{L}}, s_{\mathrm{L}}, b_{\mathrm{L}})$ are specified by 
\begin{equation}
\mathcal{L}_{\mathrm{CC}}^{(q)} = - \frac{g}{\sqrt{2}} \bar{\bi{u}}_{\mathrm{L}}\,\gamma^\mu V \bi{d}_{\mathrm{L}}W_\mu^+ + \hbox{ h.c.}\;,
\label{eqn:quarkcc}
\end{equation}
where
\begin{equation}
V \equiv U^u_{\mathrm{L}}U^{d\dagger}_{\mathrm{L}} =
\left(
\begin{array}{ccc}
{V_{ud}}  &{V_{us}}  & {V_{ub}} \\
{V_{cd}}  &{V_{cs}}  & {V_{cb}} \\
{V_{td}}  &{V_{ts}}  &  {V_{tb}}
\end{array}
\right)\;.
\label{eqn:ckmdef}
\end{equation}

The quark-mixing matrix \eref{eqn:ckmdef} is called the Cabibbo--Kobayashi-Maskawa (CKM) matrix~\cite{Cabibbo:1963yz,Kobayashi:1973fv}.
We observe significant mixing amplitudes between the first and second generations, modest mixing between the second and third, and small mixing between the first and third~\cite{Yao:2006px}:
\begin{equation}\abs{V} \equiv
\left(
\begin{array}{ccc}
\abs{V_{ud}}  & \abs{V_{us}}  &  \abs{V_{ub}} \\
\abs{V_{cd}}  & \abs{V_{cs}}  &  \abs{V_{cb}} \\
\abs{V_{td}}  & \abs{V_{ts}}  &   \abs{V_{tb}}
\end{array}
\right) =
\left(
\begin{array}{ccc}
0.974  & 0.227  &  0.004 \\
0.227  & 0.973  &  0.042 \\
0.008  & 0.042 &   0.999
\end{array}
\right)\;.
\label{eqn:ckm}
\end{equation}
The Higgs scalar is the only element of the standard model that distinguishes among the generations. In Veltman's phrase~\cite{Veltman:1997nm}, it knows something that we do not know.

In similar fashion, the charged-current interactions among the left-handed leptonic mass eigenstates $\bell_{\mathrm{L}} = (e_{\mathrm{L}}, \mu_{\mathrm{L}}, \tau_{\mathrm{L}})$ and $\bnu = (\nu_1, \nu_2, \nu_3)$ are specified by 
\begin{equation}
\mathcal{L}_{\mathrm{CC}}^{(q)} = - \frac{g}{\sqrt{2}} \,\bar{\bnu}\,\gamma^\mu \mathcal{V}^{\dagger} \bell_{\mathrm{L}}W_\mu^+ + \hbox{ h.c.}\;,
\label{eqn:leptoncc}
\end{equation}
where~\cite{Lee:1977ti}
\begin{equation}
\mathcal{V} = \left(
\begin{array}{ccc}
\mathcal{V}_{e1} & \mathcal{V}_{e2} & \mathcal{V}_{e3} \\
\mathcal{V}_{\mu1} & \mathcal{V}_{\mu2} & \mathcal{V}_{\mu3} \\
\mathcal{V}_{\tau1} & \mathcal{V}_{\tau2} & \mathcal{V}_{\tau3}
\end{array}
\right)\;.
\label{eqn:pmns}
\end{equation}
The standard form of the neutrino mixing matrix is owed to the greater familiarity of the flavor eigenstates. It is sometimes called the Pontecorvo~\cite{Pontecorvo:1957qd}--Maki-Nakagawa-Sakata~\cite{Maki:1962mu} (PMNS) matrix in tribute to neutrino-oscillation pioneers.

\begin{figure}[t!]
\begin{center}
\includegraphics[width=8.0cm]{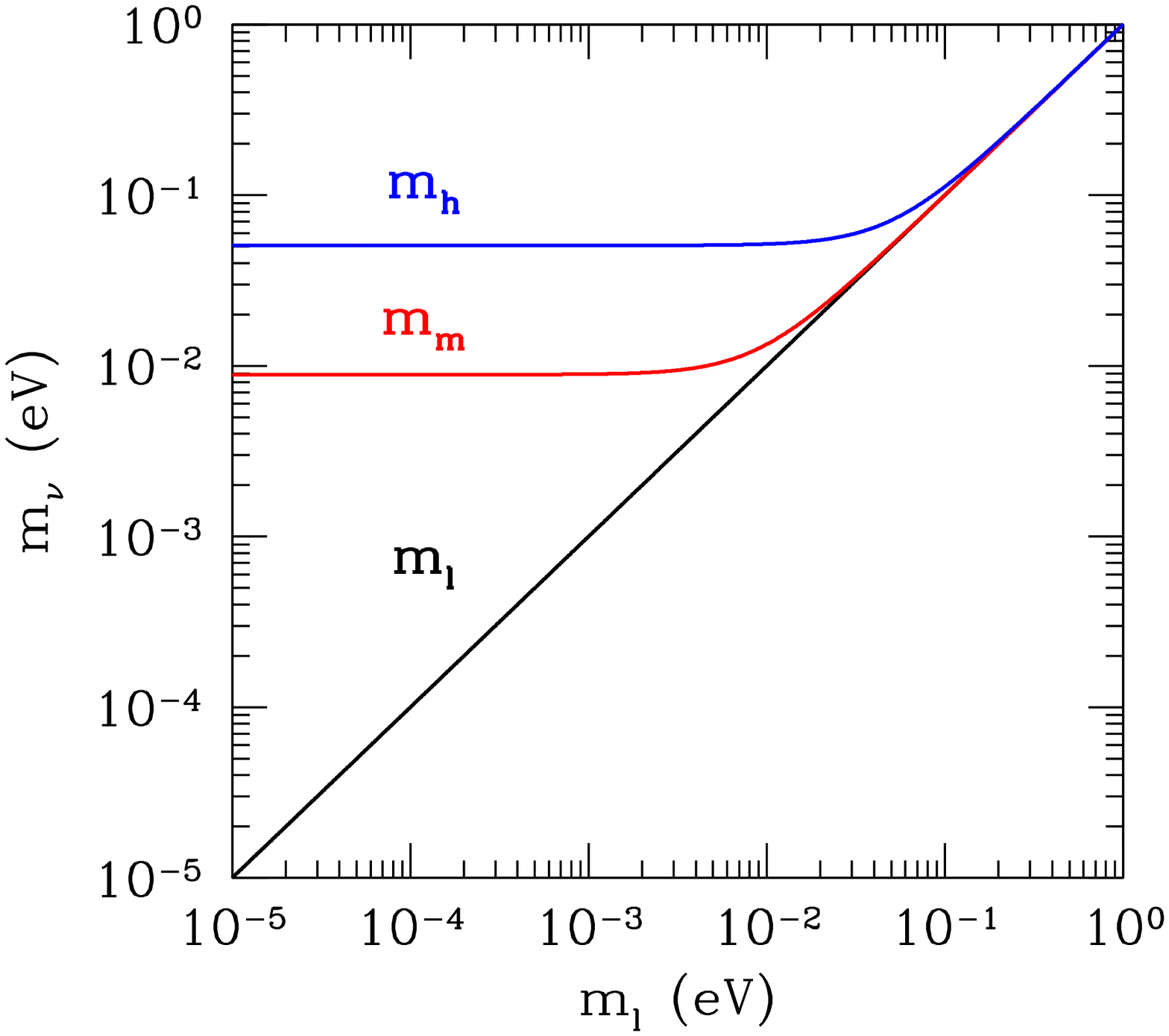}\includegraphics[width=8.0cm]{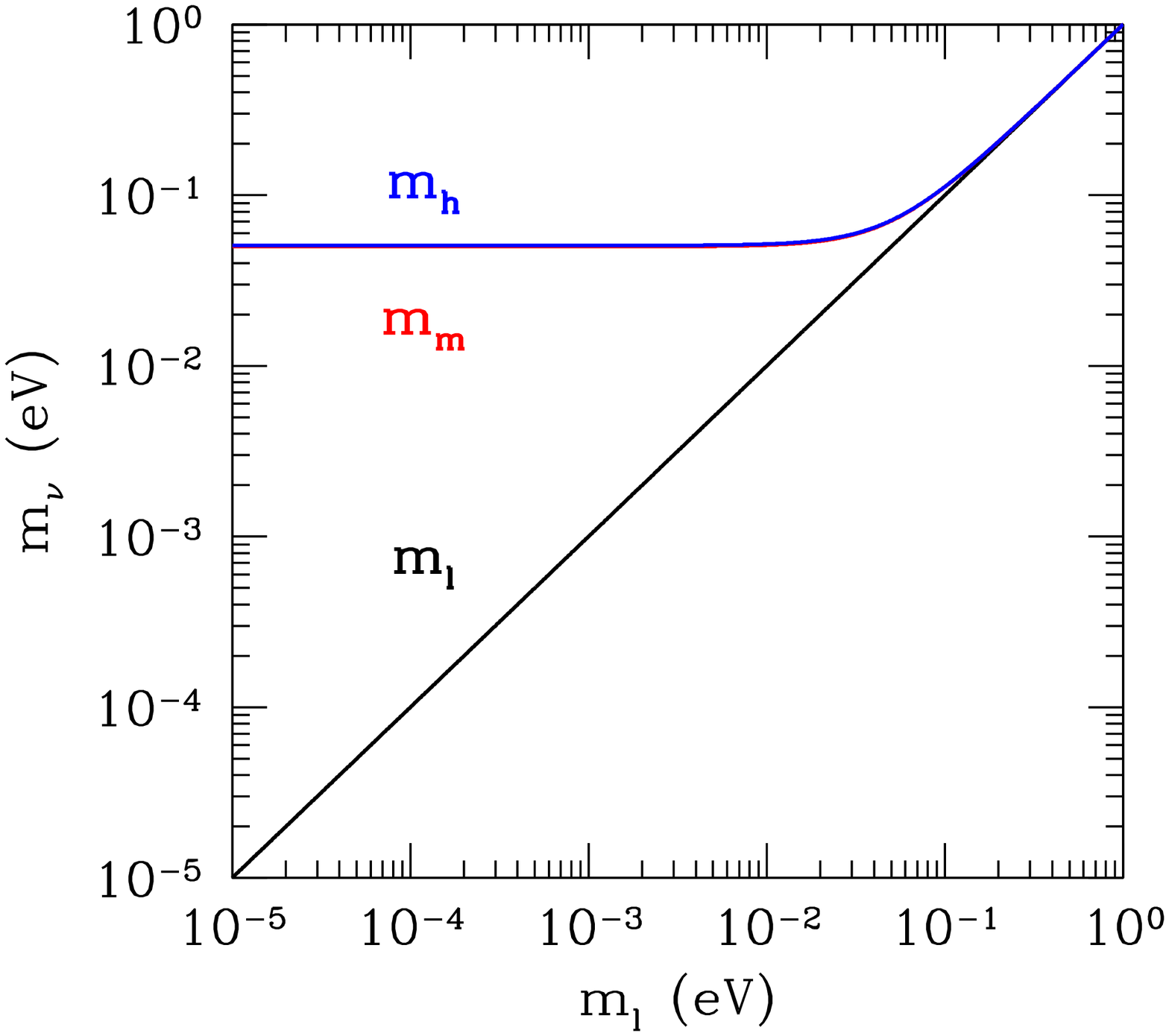}
\caption{Favored values for the light, medium, and heavy neutrino masses $m_{\ell}$, $m_{\mathrm{m}}$, $m_{\mathrm{h}}$, as functions of the lightest neutrino mass  in the three-neutrino oscillation scenario for the normal (left panel) and inverted hierarchy (right panel).
 \label{fig:neuts}} 
\end{center}
\end{figure}

By convention, $\nu_1$ and $\nu_2$ are the solar pair with $m_1 < m_2$ and $m_2^2 - m_1^2 = \Delta m^2_{\odot}$. The third mass eigenstate, $\nu_3$, is separated from $\nu_1$ and $\nu_2$ by the atmospheric mass splitting  $\Delta m^2_{\mathrm{atm}}$, but current experiments do not tell us whether it lies above (normal hierarchy) or below (inverted hierarchy) the solar doublet. We do not know the absolute scale of neutrino masses, but the effective electron-neutrino mass is constrained by kinematic measurements to be $< 2\ev$~\cite{Yao:2006px} \footnote{A plausible limit from the analysis of astronomical data is $\sum_i m_{\nu_i} < 0.62\ev$ at 95\% C.L.~\cite{Goobar:2006xz}.} We display in \fref{fig:neuts} the normal and inverted spectra that would correspond to a plausible range of values for the lightest neutrino mass, taking the current central values, $\Delta m^2_{\odot} = 7.9\times 10^{-5}\ev^2$ and $\Delta m^2_{\mathrm{atm}} = 2.5\times 10^{-3}\ev^2$~\cite{Schwetz:2006dh,Fogli:2006yq}. If the neutrino masses are Dirac masses generated (in analogy to the quark and charged-lepton masses) by $H\bar{\nu}\nu$ interactions, the Yukawa couplings are $\zeta_{\nu} \ltap 10^{-11}$. 

The observed structure of the neutrino mixing matrix differs greatly from the pattern of the quark mixing matrix \eref{eqn:ckm}. Data indicate that $\nu_3$ consists of nearly equal parts of $\nu_{\mu}$ and $\nu_{\tau}$, perhaps with a trace of $\nu_e$, while $\nu_2$ contains similar amounts of $\nu_e$, $\nu_{\mu}$, and $\nu_{\tau}$, and $\nu_1$ is rich in $\nu_e$, with approximately equal minority parts of $\nu_{\mu}$ and $\nu_{\tau}$. A recent global fit~\cite{Gonzalez-Garcia:2004jd} yields the following ranges for the magnitudes of the neutrino mixing matrix elements:
\begin{equation}
    |\mathcal{V}| =\left(
\begin{array}{ccc}
0.79-0.88&0.47-0.61&<0.20 \\
0.19-0.52& 0.42-0.73&0.58-0.82\\
0.20-0.53&0.44-0.74&0.56-0.81
\end{array}
\right)\; .
\end{equation}

What are the possible forms for neutrino mass terms? 
The familiar Dirac mass term that we have encountered for the quarks 
and charged leptons connects the left-handed and right-handed 
components of the same field,
\begin{equation}
    \mathcal{L}_{\mathrm{D}} = -D\bar{\psi}\psi = -D(\bar{\psi}_{\mathrm{L}}\psi_{\mathrm{R}}+ 
    \bar{\psi}_{\mathrm{R}}\psi_{\mathrm{L}})\;,
    \label{eq:Dmassterm}
\end{equation}
where we have used the chiral decomposition 
of a Dirac spinor,
\begin{equation}
    \psi = \half(1-\gamma_{5})\psi + \half(1 + \gamma_{5})\psi \equiv 
        \psi_{\mathrm{L}} + \psi_{\mathrm{R}}\;.
    \label{eq:chiral}
\end{equation}
The mass eigenstate is thus $\psi = \psi_{\mathrm{L}} + \psi_{\mathrm{R}}$. 

We may seek to accommodate neutrino mass in the electroweak theory by adding to the spectrum a right-handed neutrino $N_{\mathrm{R}}$ and constructing the (gauge-invariant) Dirac mass term
\begin{equation}
\mathcal{L}_{\mathrm{D}}^{(\nu)} = -\zeta_{\nu}\left[(\bar{\mathsf{L}}_{\ell}\bar{\phi})N_{\mathrm{R}} +
\bar{N}_{\mathrm{R}}(\bar{\phi}^{\dagger}\mathsf{L}_{\ell})\right]
\to -m_{\mathrm{D}}\left[\bar{\nu}_{\mathrm{L}}N_{\mathrm{R}} +
\bar{N}_{\mathrm{R}}\nu_{\mathrm{L}}\right]\;,
\label{eqn:diracnu}
\end{equation}
where $\mathsf{L}_{\ell}$ is the left-handed lepton doublet defined in \eref{eq:lleptons},
$\bar{\phi}$ is the conjugate Higgs field, $m_{\mathrm{D}} = \zeta_{\nu}v/\sqrt{2}$, and as usual $v = (G_{\mathrm{F}}\sqrt{2})^{-1/2}$. The right-handed neutrino $N_{\mathrm{R}}$ is sterile: it is an $\mathrm{SU(2)}$ singlet with weak hypercharge $Y=0$, and has no weak interactions except those induced by mixing. A Dirac mass term conserves the additive lepton number $L$ that takes on the value $+1$ for neutrinos and negatively charged leptons, and $-1$ for antineutrinos and positively charged leptons.

Because neutrinos carry neither color nor electric charge, they might---alone among the standard-model fermions---be their own antiparticles, so-called Majorana fermions. The charge conjugate of a right-handed field is left-handed,
$\psi_{\mathrm{L}}^{c} \equiv (\psi^{c})_{\mathrm{L}} =
(\psi_{\mathrm{R}})^{c}$. 
Majorana mass terms connect the left-handed and 
right-handed components of conjugate fields,
\begin{eqnarray}
-\mathcal{L}_{\mathrm{MA}} & = & A(\bar{\nu}_{\mathrm{R}}^{c}\nu_{\mathrm{L}} + 
\bar{\nu}_{\mathrm{L}}\nu_{\mathrm{R}}^{c}) = A \bar{\chi}\chi
\nonumber  \\
-\mathcal{L}_{\mathrm{MB}} & = & 
B(\bar{N}^{c}_{\mathrm{L}}N_{\mathrm{R}} + 
\bar{N}_{\mathrm{R}}N^{c}_{\mathrm{L}}) = B\bar{\omega}\omega\;.
\end{eqnarray}
The self-conjugate Majorana mass eigenstates are
\begin{eqnarray}
\chi & \equiv & \nu_{\mathrm{L}} + \nu^{c}_{\mathrm{R}} = \chi^{c}
\nonumber  \\
\omega & \equiv & N_{\mathrm{R}} + N^{c}_{\mathrm{L}} = \omega^{c} \;.
\end{eqnarray}
A Majorana fermion cannot carry any additive [$\mathrm{U(1)}$] quantum number.
The mixing of particle and antiparticle fields means that the Majorana 
mass terms correspond to processes that violate lepton number by two 
units. Accordingly, the exchange of a Majorana neutrino can mediate neutrinoless 
double beta decay, $(Z,A) \rightarrow (Z+2,A) + e^{-} + e^{-}$. 
Detecting neutrinoless double beta decay~\cite{Elliott:2002xe} would offer decisive 
evidence for the Majorana nature of the neutrino.

The mass of the active $\nu_{\mathrm{L}}$ may be generated by a Higgs triplet that acquires a vacuum expectation value~\cite{Gelmini:1980re}, or by an effective operator that involves two Higgs doublets combined to transform as a triplet~\cite{Weinberg:1979sa}.

It is interesting to consider both Dirac and Majorana terms, and specifically to examine the case in which Majorana masses corresponding to an active state $\chi$ and a sterile state $\omega$ arise from weak triplets and singlets, respectively, with masses $M_3$ and $M_1$. The neutrino mass matrix then has the form
\begin{equation}
(\begin{array}{lr}
\bar{\nu}_{\mathrm{L}} & \bar{N}^c_{\mathrm{L}}
\end{array})
\left(
\begin{array}{cc}
M_3 & m_{\mathrm{D}} \\
m_{\mathrm{D}} & M_1
\end{array}
\right)
\left(
\begin{array}{c}
\nu^c_{\mathrm{R}} \\ N_{\mathrm{R}}
\end{array}
\right)\;.
\label{eqn:mixedmass}
\end{equation}
In the highly popular seesaw limit~\cite{Minkowski:1977sc,Yanagida:1979as,Gell-Mann:1980vs,Mohapatra:1979ia,Schechter:1980gr}, with $M_3 = 0$ and $m_{\mathrm{D}} \ll M_1$, diagonalizing the mass matrix \eref{eqn:mixedmass} yields two Majorana neutrinos,
\begin{equation}
n_{1\mathrm{L}}  \approx  \nu_{\mathrm{L}} - \frac{m_{\mathrm{D}}}{M_1}N^c_{\mathrm{L}} \qquad
n_{2\mathrm{L}}  \approx  N^c_{\mathrm{L}} + \frac{m_{\mathrm{D}}}{M_1}\nu_{\mathrm{L}}\;,
\label{eqn:majopair}
\end{equation}
with masses
\begin{equation}
m_1 \approx  \frac{m_{\mathrm{D}}^2}{M_1} \ll m_{\mathrm{D}} \qquad m_2 \approx M_1\;.
\label{eqn:majomasses}
\end{equation}
The seesaw produces one very heavy ``neutrino'' and one neutrino much lighter than a typical quark or charged lepton. Many alternative explanations of the small neutrino masses have been explored in the literature~\cite{Smirnov:2004hs}, including some in which collider experiments exploring the Fermi scale could reveal the origin of neutrino masses~\cite{Chen:2006hn}.

The discovery of neutrino mass through the observation of neutrino flavor metamorphosis offers two possible paths to physics beyond the standard model. One is to accommodate right-handed sterile neutrinos in the fermion spectrum of the electroweak theory and to endow neutrinos with mass by the Higgs mechanism. To my mind, this would be a minor adjustment to the electroweak theory---correcting an oversight, we might say. Whether the small Yukawa couplings of neutrinos to the Higgs field ($\zeta_{\nu} \ltap 10^{-11}$) are qualitatively more puzzling than the factor of $3 \times 10^5$ that separates the electron and top-quark couplings is for now a question for intuition. Self-conjugate Majorana neutrinos would represent a much greater revision to the standard model because the mass of the right-handed Majorana neutrino evidently is not due to the usual Higgs mechanism. This Majorana mass might be extraordinarily large---perhaps offering a window on the unification scale for the strong, weak, and electromagnetic interactions.

\section{In Search of the Standard-Model Higgs Boson \label{sec:higgsprops}}
The evidence in hand suggests that the agent of electroweak symmetry breaking represents a novel fundamental interaction operating on the Fermi scale. \textit{We do not know what that force is.}

A leading possibility is that the agent of electroweak symmetry breaking is an elementary scalar, the Higgs boson of the electroweak standard model. An essential step toward understanding the new force that shapes our world is, therefore, to search for the Higgs boson and to explore its properties by asking
\begin{enumerate}
\item Is it there? Is there only one?
\item What are its quantum numbers?
\item Does the Higgs boson generate mass both for the electroweak gauge bosons and for the quarks and leptons?
\item How does the Higgs boson interact with itself?
\end{enumerate}
The search for the Higgs boson has been a principal goal of particle physics for many years, so search strategies have been explored in great detail~\footnote{For a useful sampler, see~\cite{Ellis:1975ap,Lee:1977eg,Vainshtein:1980ea,HHG, Djouadi:2005gi,Djouadi:2005gj,Rainwater:2007cp,DeRoeck:2004zu,Ellis:2007wa}.}. A brief profile of Higgs-boson properties and production mechanisms is in order here.

Consider first the most prominent decay modes of the standard-model Higgs boson. Decays $H \to f\bar{f}$ into fermion pairs, where $f$ occurs in $N_{\mathrm{c}}$ colors, proceed at a rate
\begin{equation}
	\Gamma(H \to f\bar{f}) = \frac{G_{F}m_{f}^{2}M_{H}}{4\pi\sqrt{2}} 
	\cdot N_{\mathrm{c}} \cdot \left( 1 - \frac{4m_{f}^{2}}{M_{H}^{2}} 
	\right)^{3/2} \; ,
	\label{eq:Higgsff}
\end{equation}
which is proportional to $N_{\mathrm{c}} m_f^2 M_{H}$ in the limit of large Higgs mass.
The partial width for decay into a $W^{+}W^{-}$ pair is
\begin{equation}
	\Gamma(H \to W^{+}W^{-}) = \frac{G_{F}M_{H}^{3}}{32\pi\sqrt{2}} 
	(1 - x)^{1/2} (4 -4x +3x^{2}) \; ,
	\label{eq:HiggsWW}
\end{equation}
where $x \equiv 4M_{W}^{2}/M_{H}^{2}$.  Similarly, the partial width 
for decay into a pair of $Z^{0}$ bosons is 
\begin{equation}
	\Gamma(H \to Z^{0}Z^{0}) = \frac{G_{F}M_{H}^{3}}{64\pi\sqrt{2}} 
	(1 - x^{\prime})^{1/2} (4 -4x^{\prime} +3x^{\prime 2}) \; ,
	\label{eq:HiggsZZ}
\end{equation}
where $x^{\prime} \equiv 4M_{Z}^{2}/M_{H}^{2}$.  The rates for decays into 
weak-boson pairs are asymptotically proportional to $M_{H}^{3}$ and 
$\cfrac{1}{2}M_{H}^{3}$, respectively.  In the final factors of \eqn{eq:HiggsWW} 
and \eqn{eq:HiggsZZ}, $2x^{2}$ and $2x^{\prime 2}$, respectively, 
arise from decays into transversely polarized gauge bosons.  The 
dominant decays for large $M_{H}$ are into pairs of longitudinally 
polarized weak bosons.  

\begin{figure}[tb]
	\centerline{\includegraphics[width=10.0cm]{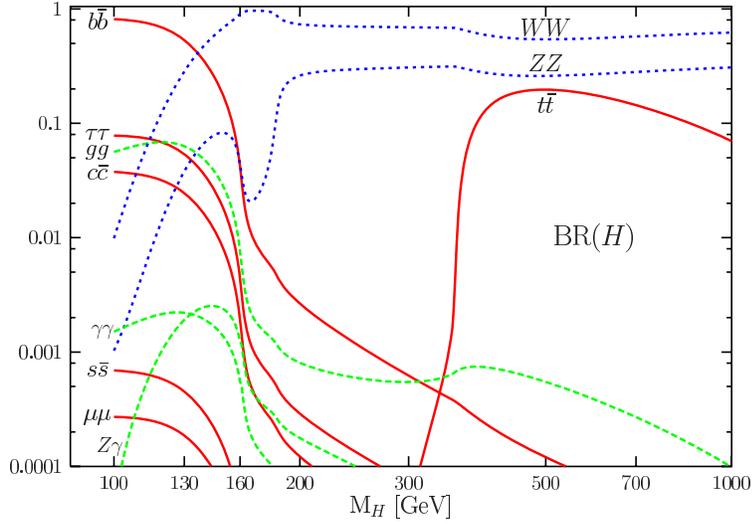}}
	\vspace*{6pt}
	\caption{Branching fractions for prominent decay modes of the standard-model Higgs boson, from~\cite{Djouadi:2005gi}.}
	\protect\label{fig:LHdk}
\end{figure}
Branching fractions for decay modes that may hold promise for the 
detection of a Higgs boson are displayed in Figure 
\ref{fig:LHdk}.  In addition to the $f\bar{f}$ and $VV$ modes that 
arise at tree level, the plot includes the $\gamma\gamma$, $Z\gamma$, and two-gluon modes that 
proceed through loop diagrams.  The rare $\gamma\gamma$ 
channel offers an important target for LHC experiments, if the Higgs boson is light, because the relatively benign backgrounds may be overcome by fine resolution.

\begin{figure}[tb]
	\centerline{\includegraphics[width=10.0cm]{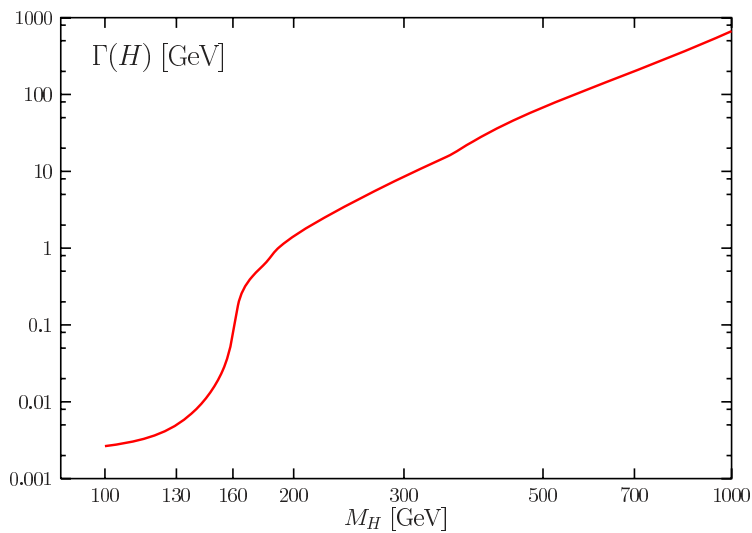}}
	\vspace*{6pt}
	\caption{Total width of the standard-model Higgs boson vs.\ mass, from~\cite{Djouadi:2005gi}.}
	\protect\label{fig:Htot}
\end{figure}

Below the $W^{+}W^{-}$ threshold, the standard-model Higgs boson is rather narrow,
with $\Gamma(H \to \mathrm{all}) \ltap 1\gev$.  Far above the threshold for decay into gauge-boson pairs, the total width is proportional to $M_{H}^{3}$.  As its mass increases toward $1\tev$, the Higgs boson becomes highly unstable, with a perturbative width approaching its mass.  It would therefore be observed as an enhanced rate, rather than a distinct resonance. The Higgs-boson total width 
is plotted as a function of $M_{H}$ in \fref{fig:Htot}.

\begin{figure}[tb]
\centerline{\includegraphics[scale=0.8]{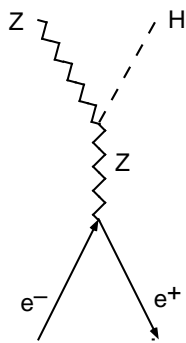}}
\vspace*{6pt}
\caption{\textit{Higgsstrahlung} mechanism for the reaction $e^+e^- \to HZ$}
\protect\label{fig:eehz}
\end{figure}

The most telling searches for the Higgs boson have been carried out at CERN's Large Electron Positron collider, LEP~\cite{Barate:2003sz,Kado:2002er}. Because the $He^+e^-$ coupling is very small, being proportional to the electron mass, the cross section for resonant Higgs-boson formation, 
$\sigma(e^{+}e^{-} \to H \to \mbox{all})$, is \textit{minute.} The small cross section sets aside a traditional strength of electron-positron annihilations---the ability to tune the collider energy to excite a resonant state. The most promising process is the radiation of a $Z$-boson and a Higgs boson from a virtual $Z$, depicted in  \fref{fig:eehz}, for which the cross section is
\begin{equation}
	\displaystyle{\sigma(e^+e^- \to HZ) = 
	\frac{\pi\alpha^{2}}{24\sqrt{s}} 
	\frac{K(K^{2}+3M_{Z}^{2})[1 + (1-4x_{W})^{2}]}{(s-M_{Z}^{2})^{2} 
	\;\;\;x_{W}^{2}(1-x_{W})^{2}}}\;,
\end{equation}
where $K$ is the c.m.\ momentum of the outgoing Higgs boson and  $x_{W}\equiv \sin^{2}\theta_{W}$.
The LEP experiments set a lower bound on the mass of the standard-model Higgs boson, $M_H > 114.4\gev$, at 95\% confidence level~\cite{Barate:2003sz,Kado:2002er}.

\begin{figure}[t!]
\begin{center}
\raisebox{1.0cm}{\includegraphics[height=5.0cm]{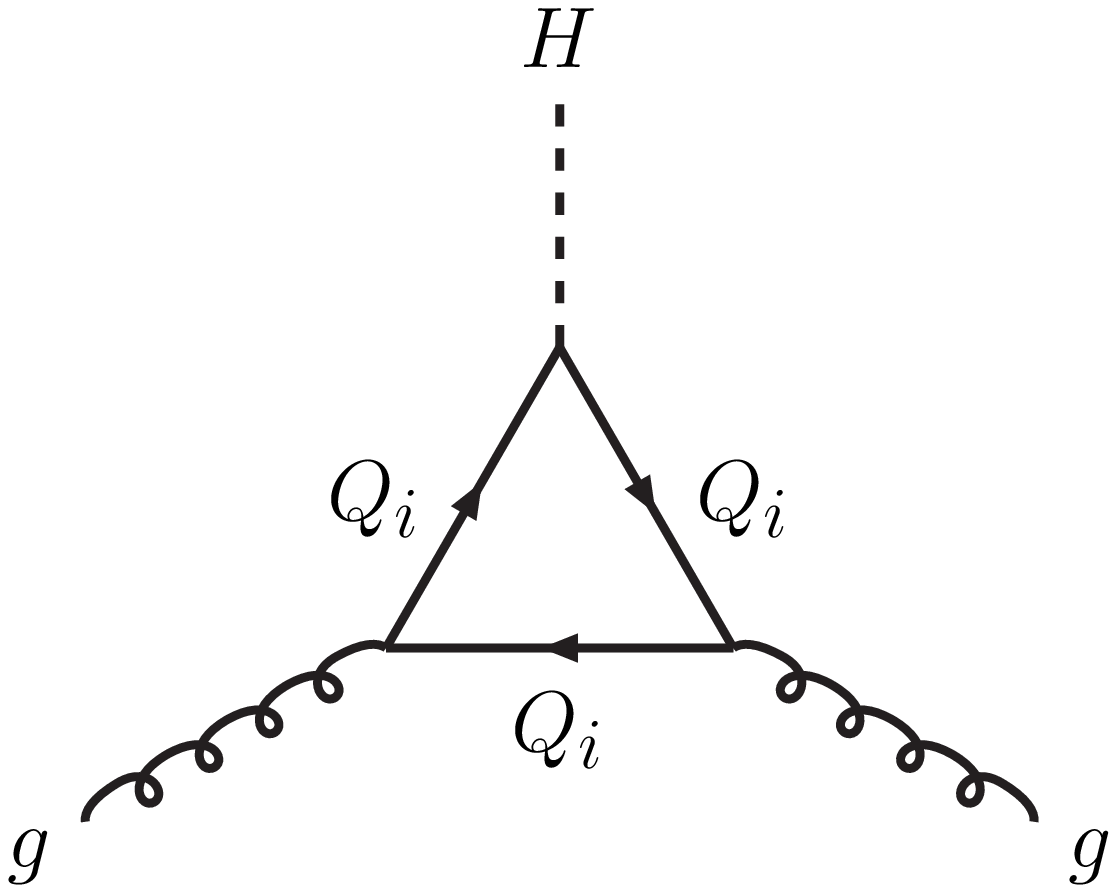}}\qquad\includegraphics[height=7.0cm]{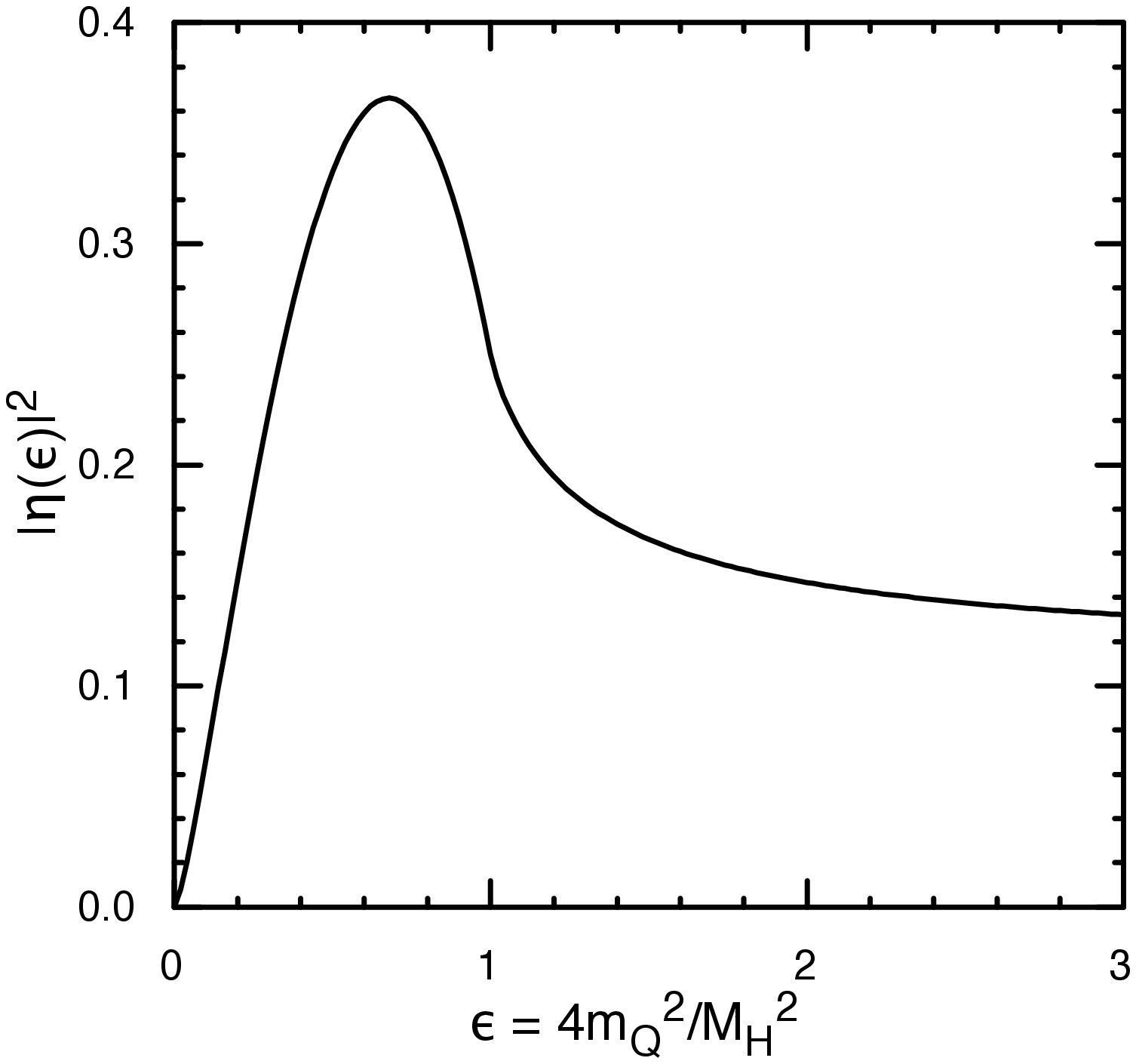}
\caption{Left panel: gluon-fusion production of a Higgs boson through a heavy-quark loop; right panel: matrix element squared for the contribution of a quark loop to $gg \to H$ (\cf \eref{eqn:toploop}) \label{fig:ggHiggs}}
\end{center}
\end{figure}

The search is now the province of the proton accelerators. The 2-TeV proton-antiproton Tevatron Collider is operating now, its integrated luminosity having surpassed $2\fb^{-1}$~\cite{tevlum}, and the 14-TeV Large Hadron Collider at CERN~\cite{lhc} will provide high-luminosity proton-proton collisions beginning in 2008 \footnote{For a prospectus on early running at the LHC, see~\cite{Gianotti:2005fm}.}. The largest cross section for Higgs production at these machines occurs in the reaction $p^{\pm}p \to H + \hbox{anything}$, which proceeds by gluon fusion through heavy-quark loops, as shown in the left-panel of  \fref{fig:ggHiggs}.
The cross section is given by
\begin{equation}
\sigma(p^{\pm}p \to H + \hbox{ anything}) = \frac{G_{\mathrm{F}}\alpha_s^2}{32\pi\sqrt{2}} \abs{\eta(\epsilon)}^2 \times (gg\hbox{ luminosity})\;,
\label{eqn:ggHiggs}
\end{equation}
where the structure of the loop diagram is captured in 
\begin{equation}
\eta(\epsilon) =\frac{\epsilon}{2}\left[1 + (\epsilon - 1)\varphi(\epsilon)\right]\;,
\label{eqn:toploop}
\end{equation}
with $\epsilon = 4m_Q^2/M_H^2$,
\begin{equation}
\varphi(\epsilon) = \left\{ \begin{array}{l}
- \arcsin^2(1/\sqrt{\epsilon}), \qquad \epsilon > 1 \\[6pt]
\case{1}{4}\left[\ln{(\zeta_+/\zeta_-)} + \rmi\pi\right]^2, \qquad \epsilon < 1\;,
\end{array}
 \right.
\label{eqn:phidef}
\end{equation}
and $\zeta_{\pm} = 1 \pm \sqrt{1 - \epsilon}$. The behavior of $\abs{\eta(\epsilon)}^2$ is plotted in the right-hand panel of  \fref{fig:ggHiggs}. Only the top-quark loop contributes significantly to $gg \to H$, in the remaining search range for $M_H$.

\begin{figure}[tb]
\begin{center}
\includegraphics[width=0.48\textwidth]{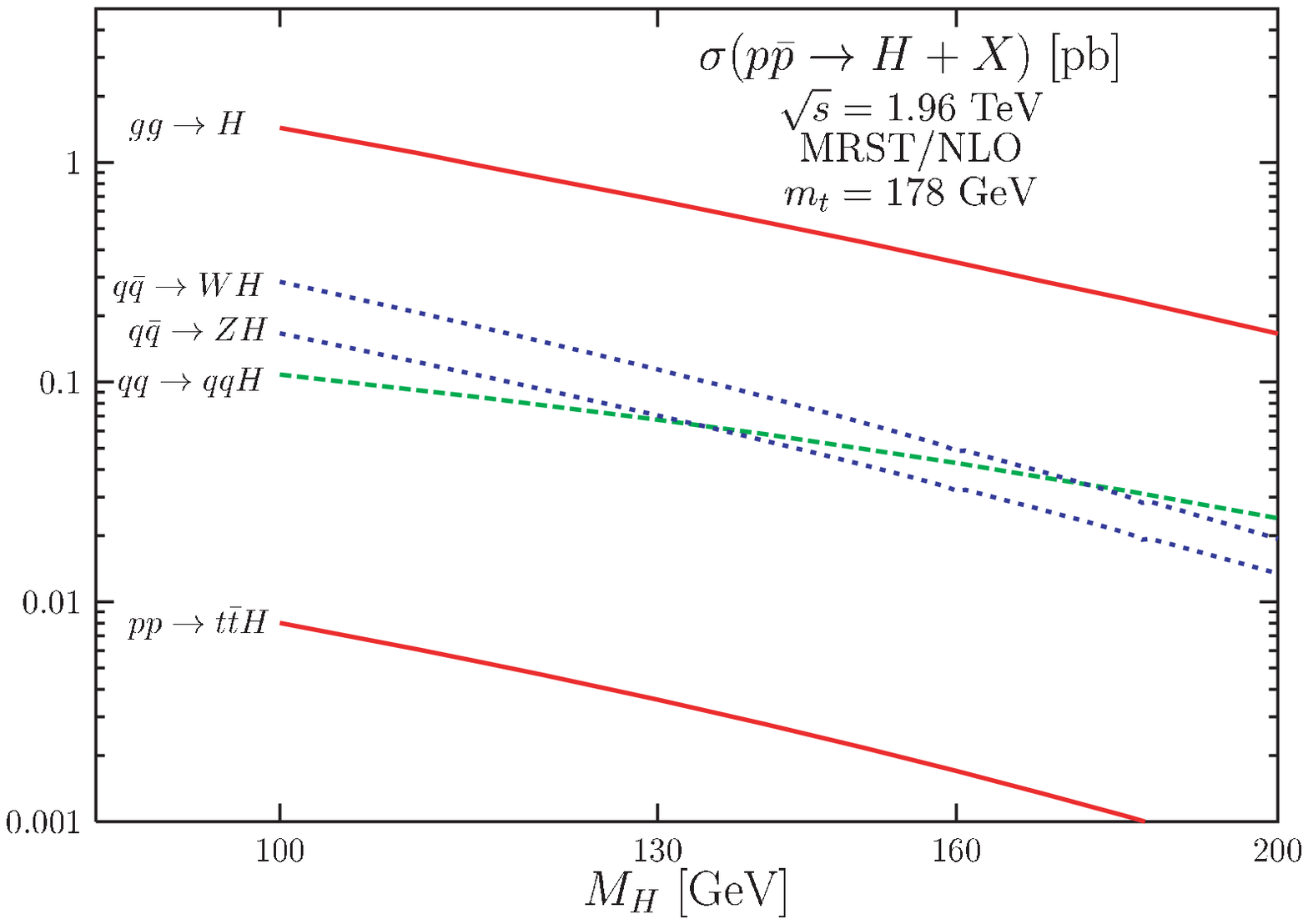}\quad\includegraphics[width=0.48\textwidth]{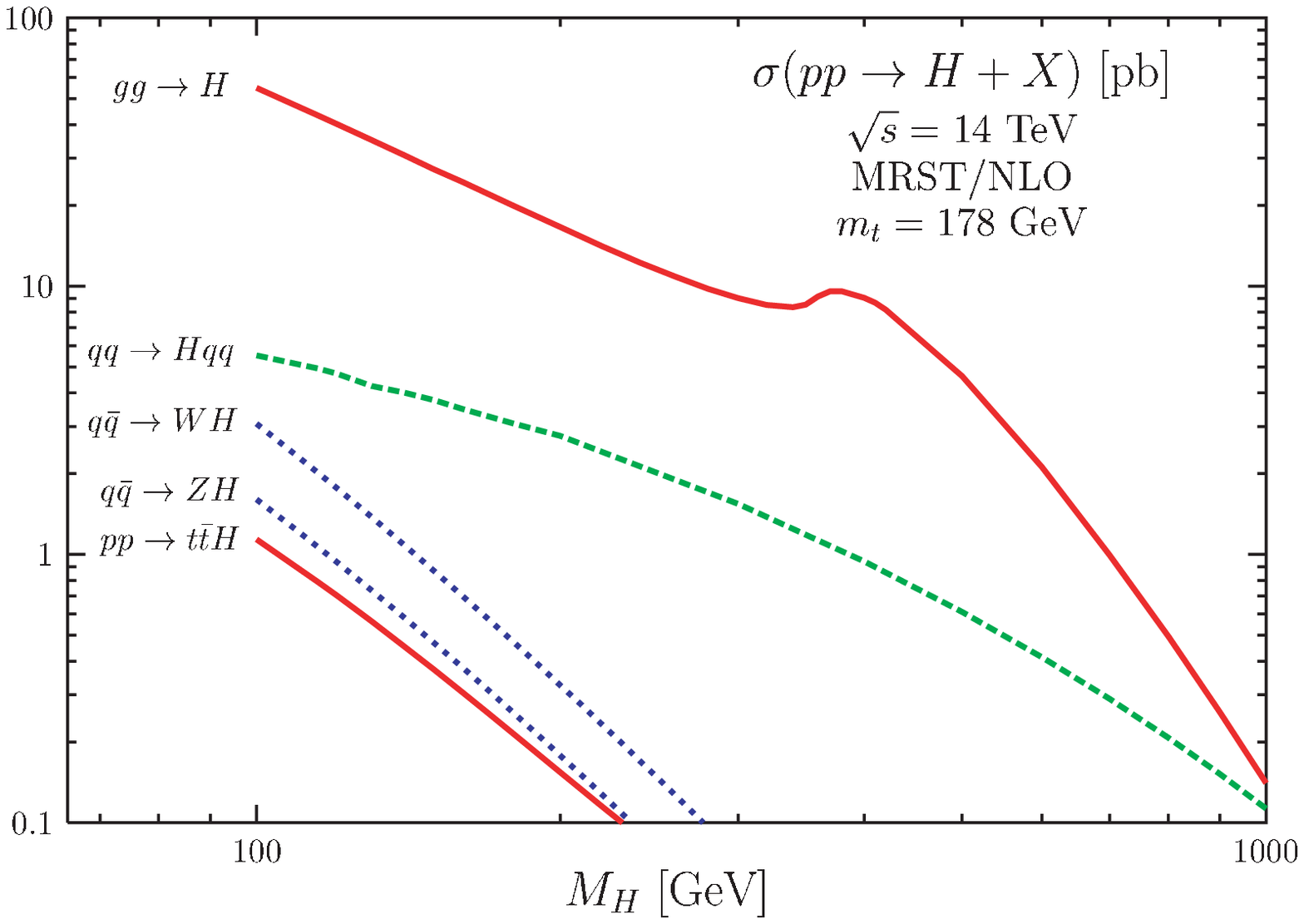}
\caption{Higgs-boson production cross sections in $\bar{p}p$ collisions at the Tevatron (left panel) and in $pp$ collisions at the LHC (right panel), from~\cite{Djouadi:2005gi} \label{fig:TeVLHCH}}
\end{center}
\end{figure}

For small Higgs-boson masses, the dominant decay is into $b\bar{b}$ pairs, but the reaction $p^{\pm}p \to H + \hbox{anything}$ followed by the decay $H \to b\bar{b}$ is swamped by QCD production of $b\bar{b}$ pairs. Consequently, experiments must rely on rare decay modes ($\tau^+\tau^-$ or $\gamma\gamma$, for example) with lower backgrounds, or resort to different production mechanisms for which specific reaction topologies reduce backgrounds. Cross sections for the principal reactions under study are shown in \fref{fig:TeVLHCH}.
[The peak of $\abs{\eta(\epsilon)}^2$ near $\epsilon = 0.68$ is reflected in the shoulder near $M_H = 400\gev$ in the $gg \to H$ cross section at $\sqrt{s} = 14\tev$.]

The production of Higgs bosons in association with electroweak gauge bosons is receiving close scrutiny at the Tevatron. Data being recorded now should give the Tevatron experiments sensitivity to raise the lower bound on the standard-model Higgs-boson mass beyond the limit set by LEP~\cite{Carena:2000yx,HiggsSens}. A twofold increase in sensitivity may begin to open the possibility of finding evidence for a light Higgs boson at the three-standard-deviation level \footnote{The status of the ongoing Tevatron searches may be tracked at~\cite{CDFHiggs,D0Higgs}.}.

At the LHC, the multipurpose detectors ATLAS~\cite{atlas} and  CMS~\cite{cms} will make a comprehensive exploration of the Fermi scale, with high sensitivity to the standard-model Higgs boson reaching to $1\tev$. Current projections suggest that a few tens of$\fb^{-1}$ will suffice for a robust discovery~\cite{Asai:2004ws}.

Once the Higgs boson is found, it will be of great interest to map its decay pattern, in order to characterize the mechanism of electroweak symmetry breaking. It is by no means guaranteed that the same agent hides electroweak symmetry and generates fermion mass. We saw in \S\ref{sec:itmatters} that chiral symmetry breaking in QCD could hide the electroweak symmetry without generating fermion masses. If it should turn out that the quarks and leptons are composite, with masses set largely by confinement energy, then the Higgs boson might couple to the masses of the constituents, not to the quark and lepton masses, with unpredictable consequences for branching fractions and production rates. Though out of theoretical fashion, quark and lepton compositeness remains a logical possibility that history reminds us not to neglect~\cite{Chivukula:2000mb}. Indeed, many extensions to the standard model significantly alter the decay pattern of the Higgs boson.  In supersymmetric models, five Higgs bosons are expected, and the branching fractions of the lightest one may be very different from those presented in \fref{fig:LHdk}~\cite{Djouadi:2005gj}. 

A Higgs-boson discovery in gluon fusion ($gg \to H$) or in the $Ht\bar{t}$ channel would argue for a nonzero coupling of the Higgs boson to top quarks, and should in time constrain the $Ht\bar{t}$ cooupling. With the large data sets the LHC will provide, it is plausible that Higgs-boson couplings can eventually be measured at levels that test the standard model and provide interesting constraints on extensions to the electroweak theory~\cite{Duhrssen:2004cv}. Precise determinations of Higgs-boson couplings is one of the strengths of the projected International Linear Collider~\cite{battaglia:163}.

We have seen the Higgs boson arise as an artifact of the mechanism we chose to hide the electroweak symmetry in \sref{sec:EWtheory}. What assurance do we have that a Higgs boson, or something very like it, will be found? It is instructive to examine the role of the Higgs boson in the cancellation of 
high-energy divergences. The most severe divergences of the individual $\nu$-, $\gamma$-, and $Z$-exchange diagrams for the production of longitudinally polarized $W^+W^-$ pairs in electron-positron collisions are tamed by a cooperation among the three diagrams of \fref{fig:eeWW}(a-c) that follows from gauge symmetry. This is not quite the end of the high-energy story. Because the electrons are massive and may therefore be found in the ``wrong'' helicity state, we must also consider a $J=0$
partial-wave amplitude, which grows as one power the c.m.\ energy.  This unacceptable high-energy behavior is precisely cancelled by the Higgs-boson graph of \fref{fig:eeWW}(d).  Something
else would have to play this role if the Higgs boson did not exist.  From the point of view of
$S$-matrix analysis, the $He\bar{e}$ coupling must be proportional to the electron mass, because the strength of ``wrong-helicity'' configurations is measured by the fermion's mass.

If the gauge symmetry were unbroken, there would be no Higgs boson, no longitudinal gauge bosons, and no extreme divergence difficulties, but we would not have a viable low-energy phenomenology
 of the weak interactions. The most severe divergences of individual diagrams are eliminated by the gauge structure of the couplings among gauge bosons and leptons. A lesser, but still potentially fatal, divergence arises because the electron has acquired mass---because of the Higgs mechanism. Spontaneous symmetry breaking provides its own cure by supplying a Higgs boson to remove the last divergence. A similar interplay and compensation must exist in any satisfactory theory. 

The sensitivity of electroweak observables to the (long unknown) mass of the top quark gave early indications for a very massive top, and the consonance of indirect and direct determinations is one element of the experimental support for the electroweak theory. Now that the top-quark mass is known to about 1.4\% from observations at the Tevatron, it becomes profitable to look beyond the quantum corrections involving top to the next most important effects, which arise from the Higgs boson. The Higgs-boson contributions are typically smaller than those from the top quark, and exhibit a more subtle (logarithmic) dependence on $M_H$ than the $m_t^2$ dependence characteristic of the top-quark contributions.

\begin{figure}[t!]
	\centerline{\includegraphics[width=10.0cm]{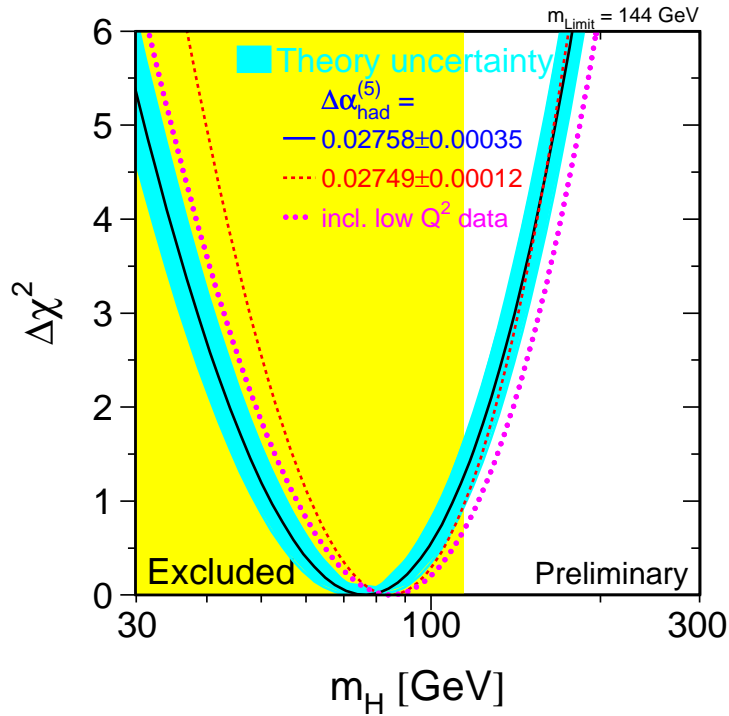}}
	\vspace*{6pt}
	\caption{$\Delta\chi^2 = \chi^2 - \chi_{\mathrm{min}}^2$ from a fit to a universe of electroweak measurements as a function of the standard-model Higgs-boson mass. The solid line is the result of the fit.The blue band represents an estimate of the theoretical uncertainty due to missing higher-order corrections. The vertical yellow band shows the 95\% CL lower bound on $M_H$ from direct searches at LEP. The dashed curve shows the sensitivity to a change in the evaluation of $\alpha_{\mathrm{em}}(M_Z^2)$. (From the LEP Electroweak Working Group~\cite{lepewwg}.)}
	\protect\label{fig:blueband}
\end{figure}

 \Fref{fig:blueband} shows how the goodness of the LEP Electroweak Working Group's Winter 2007 
global fit depends upon $M_H$ \footnote{See~\cite{Rosner:2001zy} for an introduction to global analyses.}. That the fit is improved by the inclusion of Higgs-boson effects  \footnote{See~\cite{Chanowitz:1999se} for an estimate of how strong nonresonant $WW$ scattering would modify radiative corrections.} does
not constitute proof that the standard-model Higgs boson exists, but does show that the consistency of a standard-model analysis favors a light Higgs boson. The precision electroweak measurements on their own argue for $M_H \ltap 144\gev$, a one-sided 95\% confidence level limit derived from $\Delta\chi^2 = 2.7$ for the blue band in \fref{fig:blueband}. Imposing the exclusion $M_H > 114.4\gev$ from the LEP searches leads to an upper bound of  $M_H \ltap 182\gev$~\cite{lepewwg}.


The mass favored by the global fit, $M_H = 76^{+33}_{-24}\gev$, lies in the region excluded by direct searches. Chanowitz~\cite{Chanowitz:2002cd,Chanow07} has cautioned that the values of $M_H$ preferred by fits to different observables are not entirely consistent. In particular, the forward-backward asymmetry in $e^+ e^- \to b\bar{b}$ on the $Z$ resonance is best reproduced with $M_H \approx 400\gev$. This is the observable most discrepant ($\approx 2.9\sigma$) with the overall fit; omitting it would improve the global fit, but lead to a very small Higgs-boson mass that would coexist uncomfortably with the LEP exclusion. Whether this tension is a harbinger of new physics or merely a statistical fluctuation is a tantalizing question.

By demanding consistency of the electroweak theory as a quantum field theory, we can establish bounds on the Higgs boson mass, and uncover another reason to expect that 
discoveries will not end with the Higgs boson.  Scalar field theories 
make sense on all energy scales only if they are noninteracting, or 
``trivial''~\cite{Wilson:1973jj}.  The vacuum of quantum field theory is a dielectric 
medium that screens charge.  Accordingly, the effective charge is a 
function of the distance or, equivalently, of the energy scale.  This is 
the famous phenomenon of the running coupling constant.

In $\lambda\phi^4$ theory (compare the interaction term in the Higgs 
potential), it is easy to calculate the variation of the coupling 
constant $\lambda$ in perturbation theory by summing quantum corrections given by bubble graphs \footnote{Lattice field theory allows us to treat the strong-coupling regime beyond the small-$\lambda$ realm in which perturbation theory can be trusted.}. The coupling constant $\lambda(\kappa)$ on a physical scale $\kappa$ 
is related to the coupling constant on a higher scale $\Lambda$ by
\begin{equation}
\frac{1}{\lambda(\kappa)} = \frac{1}{\lambda(\Lambda)} + 
\frac{3}{2\pi^2}\log{\left(\Lambda/\kappa\right)}\;\;.
\label{rng}
\end{equation}

In order for the Higgs potential to be stable (\ie, for the energy of the 
vacuum state not to race off to $-\infty$), $\lambda(\Lambda)$ must not 
be negative.  Applied to \eqn{rng}, this condition leads to an inequality,
\begin{equation}
\frac{1}{\lambda(\kappa)} \ge 
\frac{3}{2\pi^2}\log{\left(\Lambda/\kappa\right)}\;\;,
\end{equation}
that implies an {\em upper bound},
\begin{equation}
\lambda(\kappa) \le 
2\pi^2/3\log{\left(\Lambda/\kappa\right)}\;\;,
\label{upb}
\end{equation}
on the coupling strength at the physical scale $\kappa$.
If  the theory is to make sense to arbitrarily high energies---or 
short distances---we must consider the limit $\Lambda\rightarrow\infty$ 
while holding $\kappa$ fixed at some reasonable physical scale;  
 the bound \eqn{upb} then forces $\lambda(\kappa)$ to zero.  The scalar field 
theory has become free field theory; in theorist's jargon, it is trivial.

Rearranging and exponentiating both sides of  \eqn{upb} gives the condition
\begin{equation}
\Lambda \le \kappa \exp{\left(\frac{2\pi^2}{3\lambda(\kappa)}\right)}\;\;,
\end{equation}
from which we can infer a limit on the Higgs-boson mass.
Choosing the physical scale as $\kappa=M_H$, using the definition $M_H^2 = 2\lambda(M_H)v^2$,
we find that
\begin{equation}
\Lambda \le \Lambda^{\star} \equiv M_H\exp{\left(\frac{4\pi^2v^2}{3M_H^2}\right)}\;\;.
\label{eqn:maxlambda}
\end{equation}
For any given Higgs-boson mass, we can identify a maximum energy scale 
$\Lambda^\star$ at which the theory ceases to make sense.  The 
description of the Higgs boson as an elementary scalar is at best an 
effective theory, valid over a finite range of energies.

\begin{figure}[tb]
	\centerline{\includegraphics[width=10.0cm]{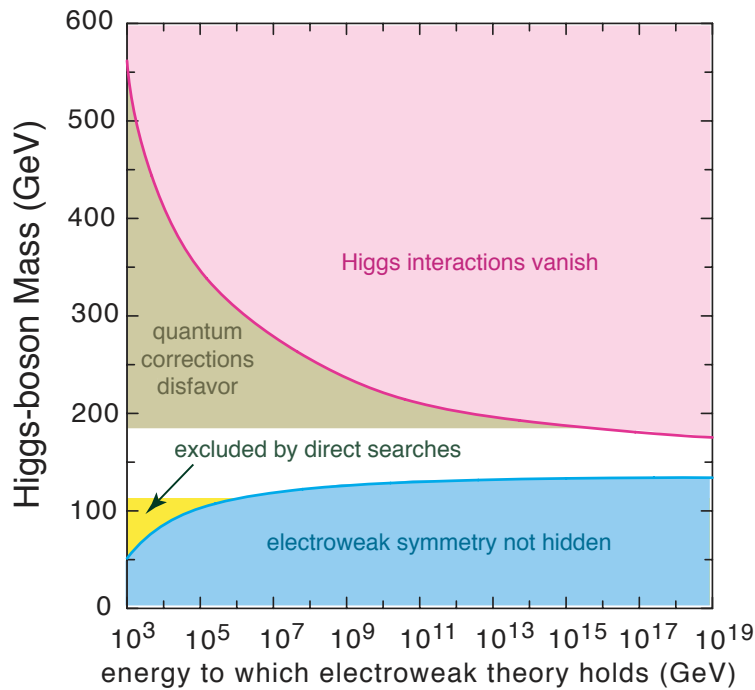}}
	\vspace*{6pt}
	\caption{Bounds on the standard-model Higgs-boson mass that follow from 
	requirements that the electroweak theory be consistent up to the 
	energy $\Lambda$.  The upper bound follows from triviality 
	conditions; the lower bound follows from the requirement that $V(v) < 
	V(0)$.  Also shown is the range of masses permitted at the 95\%\ 
	confidence level by precision measurements and direct searches.}
	\protect\label{fig:constraints}
\end{figure}

A lower bound on $M_H$ is obtained by 
computing~\cite{Linde:1975sw,Weinberg:1976pe,Hung:1979dn,Sher:1988mj,Schrempp:1996fb,Altarelli:1994rb} quantum corrections to the classical potential
\eqn{SSBpot} and requiring that $\vev{\phi} = v/\!\sqrt{2}$ be an absolute minimum of the Higgs 
potential. The upper and lower bounds plotted in \fref{fig:constraints} are the results of 
full two-loop calculations~\cite{Casas:1994us,Hambye:1996wb}.  There I have also 
indicated the upper bound on $M_{H}$ derived from precision 
electroweak measurements in the framework of the standard electroweak 
theory, as well as the lower limit from direct searches for the Higgs boson. Evidently the theory can be self-consistent up to very high energies, provided that the Higgs boson is relatively light.  For the electroweak theory to make sense all the 
way up to a unification scale $\Lambda^\star = 10^{16}~\hbox{GeV}$, 
the Higgs-boson mass must lie in the interval $134\gev \ltap M_{H}
\ltap 177 \gev$.  If $M_H$ is not within this chimney, the electroweak theory is incomplete; it is an effective theory that will be subsumed in a more comprehensive description.

This perturbative analysis leading to \eref{eqn:maxlambda} breaks down when the Higgs-boson mass 
approaches $1\tev$ and the interactions become strong.  
Lattice analyses~\cite{Heller:1993yv} indicate that, for the theory to describe 
physics to an accuracy of a few percent up 
to a few TeV, the mass of the Higgs boson can be no more than about 
$710\pm 60\gev$.  If the elementary Higgs boson takes on the largest mass allowed by 
perturbative unitarity arguments, the electroweak theory lives
on the brink of instability.

\begin{figure}[tb]
	\centerline{\includegraphics[width=12.0cm]{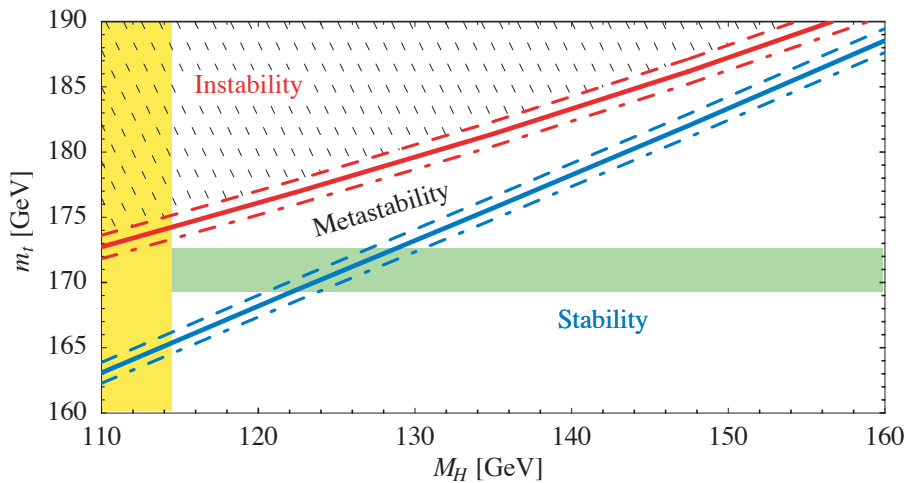}}
	\vspace*{6pt}
	\caption{Metastability region of the standard-model vacuum in the $(M_H,m_t)$ plane, from~\cite{Isidori:2001bm}. The solid curves are calculated for the choice $\alpha_s(m_Z)=0.118$. Dashed and dot-dashed curves show the effects of varying $\alpha_s(m_Z)$ by $\pm 0.002$. The horizontal green band indicates the measured top-quark mass, $m_t = 170.9 \pm 1.8\gev$. The yellow band at the left shows the region of Higgs-boson masses excluded by searches at LEP,  $M_H \ge 114.4\gev$. [After~\cite{Isidori:2001bm}.]} 	
	\protect\label{fig:metastab}
\end{figure}

The condition of absolute vacuum stability that leads to the lower bound on $M_H$ displayed in \fref{fig:constraints} is more stringent than is required by observational evidence. It would suffice to ensure that the presumed ground state of the electroweak theory has survived quantum fluctuations until now, so that the mean time to tunnel from our vacuum to a deeper vacuum at large values of $\abs{\phi}$ exceeds the age of the Universe, $T_{\mathrm{U}} \approx 13.7\hbox{ Gy}$~\cite{Yao:2006px}.  \Fref{fig:metastab} shows the outcome of a renormalization-group-improved one-loop calculation of the tunneling probability at zero temperature~\cite{Isidori:2001bm, Froggatt:2001pa,Einhorn:2007rv}. For $M_H = 115\gev$, the Higgs potential develops an instability below the Planck scale for values of the top-quark mass $m_t \gtap 166\gev$, but the electroweak vacuum's lifetime exceeds $T_{\mathrm{U}}$ so long as $m_t \ltap 175\gev$. Should $(M_H,m_t)$ settle in the metastable zone, we would have a provocative hint for new physics below the Planck scale.

\section{More New Physics on the Fermi Scale?}
\begin{figure}[tb]
\centerline{\includegraphics[width=10.0cm]{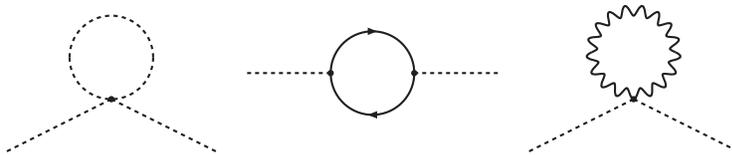}}
\vspace*{6pt}
\caption{Examples of loop diagrams that enter in the computation of quantum corrections to the Higgs-boson mass. The dashed lines represent the Higgs boson, solid lines with arrows represent fermions and antifermions, and wavy lines stand for gauge bosons.}
\protect\label{fig:mechancete}
\end{figure}

Contemplating quantum corrections to the Higgs-boson mass leads to the inference that more new phenomena may appear at energies near $1\tev$. Suppose that, in keeping with the unitarity argument reviewed in \sref{sec:import}, a Higgs boson is found with mass $M_H \ltap 1\tev$. 
How is the scale of electroweak symmetry breaking maintained in the presence 
of quantum corrections?   Beyond the classical approximation, scalar mass parameters receive 
quantum corrections from loops that contain particles of spins 
$J=0, \cfrac{1}{2}$, and $1$, as indicated schematically in \fref{fig:mechancete}.
The loop integrals that determine the running mass lead potentially to divergences, which we may characterize by
\begin{equation}
	M_H^2(p^2) = M_H^2(\Lambda^2) + Cg^2\int^{\Lambda^2}_{p^2}dk^2 
	+ \cdots \;,
	\label{longint}
\end{equation}
where $\Lambda$ defines a reference scale at which the value of 
$M_H^2$ is known, $g$ is the coupling constant of the theory, and the 
coefficient $C$ is calculable in any particular theory.  
The loop integrals appear to be quadratically divergent, $\propto \Lambda^2$. 
In order for the mass shifts 
induced by quantum corrections to remain under control, either something must limit the range of integration, or new physics must otherwise intervene to damp the integrand.

In the absence of new physics, the reference scale $\Lambda$ would naturally be large.
If the fundamental interactions are described by quantum
chromodynamics and the electroweak theory, then a 
natural reference scale is the Planck mass,
$\Lambda \sim M_{\rm Planck}  = 
	\left({\hbar c}/{G_{\mathrm{Newton}}}\right)^{1/2} \approx 1.2 
	\times 10^{19}\gev$.
In a unified theory of the strong, weak, and electromagnetic 
interactions, a natural scale is the unification scale,
$\Lambda \sim M_U \approx 10^{15}\hbox{ - }10^{16}\gev$.
 Both estimates are very large compared to the electroweak scale, and so imply a very long range of integration.  The challenge of preserving widely separated electroweak and reference scales in the presence of quantum corrections is known as the \textit{hierarchy problem.} Unless we suppose that $M_H^2(\Lambda^2)$ and the quantum corrections are finely tuned to yield $M_H^2(p^2) \ltap (1\tev)^2$, some new physics must intervene at an energy of approximately $1\tev$ to bring the integral in \Eqn{longint} under control \footnote{The analysis is more subtle if the reference scale $\Lambda$ does not represent a physically meaningful cutoff. In that event, evaluating the loop integral by dimensional regulation~\cite{tHooft:1972fi}  yields no $\Lambda^2$ piece. However, the scale of $M_H^2$ is naturally set by the heaviest particles to which the Higgs boson couples---directly, or even through gauge bosons, so a hierarchy problem persists. For a careful treatment, see \S1 of~\cite{Martin:1997ns}.}. 
 
A fine-tuning problem may be seen to arise even when the scale $\Lambda$ is not extremely large. What has been called the ``LEP Paradox''~\cite{Barbieri:2000gf,Burdman:2006tz} refers to a tension within the precise measurements of electroweak observables carried out at LEP and elsewhere. On the one hand, the global fits summarized in \fref{fig:blueband} point to a light standard-model Higgs boson. On the other hand, a straightforward effective-operator analysis of possible beyond-the-standard-model contributions to the same observables gives no hint of any new physics---of the kind needed to resolve the hierarchy problem---below about $5\tev$.
\begin{figure}[t!]
\begin{center}
\includegraphics[width=10.0cm]{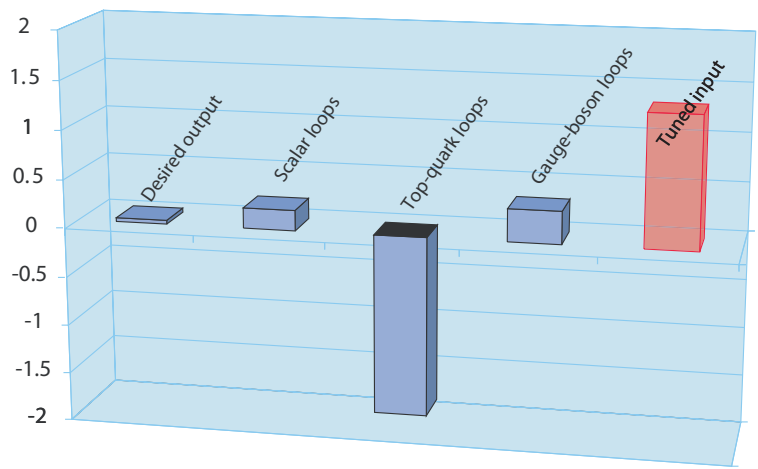}
\caption{Relative contributions to $\Delta M_H^2$ for $\Lambda = 5\tev$ \label{fig:finetune2}}
\end{center}
\end{figure}
\Fref{fig:finetune2} shows that even with a cutoff $\Lambda = 5\tev$, a careful balancing act is required to maintain a small Higgs-boson mass in the face of quantum corrections. The chief culprit is the large contribution from the top-quark loop. We are left to ask what enforces the balance, or how we might be misreading the data.

Let us review the argument for the hierarchy problem: The unitarity argument showed that new physics must be present on the \onetev, either in the form of a Higgs boson, or other new phenomena. But a low-mass Higgs boson is imperiled by quantum corrections. New physics not far above the \onetev\ could bring the reference scale $\Lambda$ low enough to mitigate the threat. If the reference scale is indeed very large, then either various contributions to the Higgs-boson mass must be precariously balanced or new physics must control the contribution of the integral in \Eqn{longint}. We do not have a proof that Nature is not fine tuned, but I think it highly likely that \textit{both a Higgs boson and other new phenomena} are to be found near the \onetev.

A new symmetry, not present in the standard model, could resolve the hierarchy problem.
  Exploiting the fact 
that fermion loops contribute with an overall minus sign relative to boson loops (because of 
Fermi statistics), \textit{supersymmetry}~\cite{Martin:1997ns,Djouadi:2005gj,Pape:2006ar} balances the contributions of fermion and boson loops \footnote{``Little Higgs" models~\cite{Schmaltz:2005ky} and ``twin Higgs'' models~\cite{Chacko:2005pe} employ  different conspiracies of contributions to defer the hierarchy problem to about $10\tev$.}.  In unbroken supersymmetry, the 
masses of bosons are degenerate with those of their fermion 
counterparts, so the cancellation is exact. If supersymmetry is present in our world, it must be broken.
The contribution of the integrals may still be acceptably small if the 
fermion-boson mass splittings $\Delta M$ are not too large.  The 
condition that $g^2\Delta M^2$ be ``small enough'' leads to the 
requirement that superpartner masses be less than about 
$1\tev$. It is provocative to note that, with superpartners at $\mathcal{O}(1\tev)$, the $\mathrm{SU(3)_c}\otimes \ewgg$ coupling constants run to a common value at a unification scale of about $10^{16}\gev$~ \cite{Amaldi:1991cn}.

Theories of dynamical symmetry breaking (\cf \sref{sec:dsb2}) offer a  second solution to the problem of the enormous range of integration in 
\eqn{longint}. In technicolor models, the Higgs boson is composite, and its internal structure comes into play  on the scale of its binding, $\Lambda_{\mathrm{TC}} \simeq 
\mathcal{O}(1~{\rm TeV})$. The integrand is damped, the effective range of integration is cut off, and 
mass shifts are under control.

Dark matter offers one more independent indication that new phenomena should be present on the Fermi scale. An appealing interpretation of the evidence that dark matter makes up roughly one-quarter of the energy density of the Universe~\cite{Spergel:2006hy} is that dark matter consists of  thermal relics of the big bang: stable---or exceedingly long-lived---neutral particles. If the particle has couplings of weak-interaction strength, then generically the observed dark-matter density results if the mass of the dark-matter particle lies between approximately $100\gev$ and $1\tev$~\cite{Bertone:2004pz}. 
Whether based on extra dimensions, new strong dynamics, or supersymmetry, scenarios to extend the electroweak theory and resolve the hierarchy problem typically entail dark-matter candidates on the Fermi scale \footnote{Other dark-matter candidates, notably the axions~\cite{Weinberg:1977ma,Wilczek:1977pj} implicated in a possible solution~\cite{Peccei:1977hh,Peccei:1977ur} to the strong \textsf{CP} problem, do not select the \onetev. For recent reviews, see~\cite{Yao:2006px,Raffelt:2006rj}.}.

\section{Another look at dynamical symmetry breaking \label{sec:dsb2}}
Quantum chromodynamics offers a suggestive model for the idea of dynamical symmetry breaking, even if it does not account itself satisfactorily for the observed electroweak symmetry breaking.
The minimal \textit{technicolor} model~\cite{Weinberg:1976gm,Susskind:1979ms}  
transcribes the same ideas from QCD to a new setting.  The 
technicolor gauge group is taken to be $\mathrm{SU}(N)_{\mathrm{TC}}$ (usually 
$\mathrm{SU(4)}_{\mathrm{TC}}$), 
so the gauge interactions of the theory are generated by
\begin{equation}
	\mathrm{SU(4)_{\mathrm{TC}}\otimes SU(3)_c \otimes SU(2)_L \otimes U(1)_\mathit{Y}}\; .
\end{equation}
The technifermions are a chiral doublet of massless color singlets
\begin{equation}
\begin{array}{cc}
	\left( \begin{array}{c} U \\ D \end{array} \right)_\mathrm{L} & U_\mathrm{R}, \;
D_\mathrm{R} \; .
\end{array}
\end{equation}
With the electric charge assignments $Q(U)=\frac{1}{2}$ and
$Q(D)=-\frac{1}{2}$, the  
theory is free of electroweak anomalies. The ordinary fermions are all 
technicolor singlets. 

In analogy with our discussion of chiral symmetry breaking in QCD, we 
assume that the chiral symmetry of the massless technifermions is broken,
\begin{equation}
	\mathrm{SU(2)_L\otimes SU(2)_R\otimes U(1)_V\to SU(2)_V\otimes U(1)_V}\; .
\end{equation}
Three would-be Goldstone bosons, the  technipions $\pi_{\mathrm{T}}^+,  \pi_{\mathrm{T}}^0,  \pi_{\mathrm{T}}^-$, emerge,
for which we are free to {\em choose} the technipion decay constant as
\begin{equation}
	F_\pi = \left(G_{\mathrm{F}}\sqrt{2}\right)^{-1/2} = 246\gev\; , \label{FPI}
\end{equation}
which amounts to choosing the scale on which technicolor becomes strong.
When the electroweak interactions are turned on, the technipions become the 
longitudinal components of the intermediate bosons, which acquire canonical standard-model masses
\begin{equation}
\renewcommand\arraystretch{1.5}
\begin{array}{ccccc}
	M_W^2 & = & g^2F_\pi^2/4 & = & 
{\displaystyle \frac{\pi\alpha}{G_{\mathrm{F}}\sqrt{2}\sin^2\theta_W}} \\
	M_Z^2 & = & \left(g^2+g^{\prime 2}\right)F_\pi^2/4 & = & 
M_W^2/\cos^2\theta_W\; ,
\end{array} 
\renewcommand\arraystretch{1}
\end{equation}
thanks to our choice \eref{FPI} of the technipion decay constant \footnote{An alternative variation on the QCD theme posits that QCD-induced chiral-symmetry breaking in exotic (color $\mathbf{6}, \mathbf{8}, \mathbf{10,}$ \ldots) quark sectors may drive electroweak symmetry breaking~\cite{PhysRevD.21.2425}.}.

Technicolor shows how the generation of intermediate boson masses 
could arise without fundamental scalars or unnatural adjustments of 
parameters.  It thus provides an elegant solution to the naturalness 
problem of the standard model.  However, it 
offers no explanation for the origin of quark and lepton masses, 
because no Yukawa couplings are generated between Higgs fields and 
quarks or leptons.  Consequently, technicolor serves as a reminder 
that \textit{particle physics confronts two problems of mass:} explaining the masses of the 
gauge bosons, which demands an understanding of electroweak symmetry 
breaking; and accounting for the quark and lepton masses, which 
requires not only an understanding of electroweak symmetry breaking 
but also a theory of the Yukawa couplings that set the scale of 
fermion masses in the standard model. 

We can be confident that the 
origin of gauge-boson masses will be understood on the Fermi scale.  
We do not know where we will decode the pattern of the Yukawa 
couplings. In extended technicolor models~\cite{Dimopoulos:1979es,Eichten:1979ah,Appelquist:2003hn}, separate gauge interactions hide the electroweak symmetry and communicate the broken symmetry to the quarks and leptons. Specific implementations of these ideas face phenomenological challenges pertaining to flavor-changing neutral currents, the large top-quark mass, and precision electroweak measurements, but the idea of dynamical symmetry breaking remains an important alternative to the standard elementary scalar~\cite{Hill:2002ap,Lane:2002wv}.

\section{The Vacuum Energy Puzzle} 
The cosmological constant problem---why empty space is so nearly massless---is one of the great mysteries of science~\cite{RevModPhys.61.1,Abbott:1988nx}. It is the reason why gravity has weighed on the minds of electroweak theorists~\cite{Linde:1974at,Veltman:1997nm,Veltman:1974au}, despite the utterly negligible role that gravity plays in particle reactions.

At the vacuum expectation value $\vev{\phi}$ of the Higgs field, the (position-independent) value of the Higgs potential is 
\begin{equation}
    V(\vev{\phi^{\dagger}\phi}) = \frac{\mu^{2}v^{2}}{4} = 
    - \frac{\abs{\lambda}v^{4}}{4} < 0.
    \label{minpot}
\end{equation}
Identifying $M_{H}^{2} = -2\mu^{2}$, we see that the Higgs potential 
contributes a field-independent constant term,
\begin{equation}
    \varrho_{H} \equiv \frac{M_{H}^{2}v^{2}}{8},
    \label{eq:rhoH}
\end{equation}
which plays the role of a vacuum energy density $\varrho_{H}$ in the Lagrangian.
In the context of gravitation, this is equivalent to adding a cosmological constant 
term, $\Lambda = (8\pi G_{\mathrm{Newton}}/c^{4})\varrho_{H}$, to Einstein's equation~\cite{Peebles:2002gy}.  

Recent observations of the accelerating expansion of the Universe~\cite{Riess:1998cb,Perlmutter:1998np} raise the intriguing possibility that the cosmological constant may be different from zero, but the essential fact is that the observed vacuum energy density must be very small indeed~\cite{Yao:2006px},
\begin{equation}
    \varrho_{\mathrm{vac}} \ltap 10^{-46}\gev^{4} \approx (\hbox{a few meV})^4\; .
    \label{eq:rhovaclim}
\end{equation}
Therein lies the puzzle: if we take
$v = (G_{\mathrm{F}}\sqrt{2})^{-\frac{1}{2}}  \approx 246\gev$  
and insert the current experimental lower bound~\cite{Barate:2003sz} 
$M_{H} \gtap 114.4\gev$ into \eqn{eq:rhoH}, we find that the Higgs field's
contribution to the vacuum energy density is
\begin{equation}
    \varrho_{H} \gtap  10^{8}\gev^{4},
    \label{eq:rhoHval}
\end{equation}
some 54 orders of magnitude larger than the upper bound inferred from 
the cosmological constant. This mismatch has been a source of dull headaches for more than three decades.

The problem is still more serious in a unified theory of the strong, 
weak, and electromagnetic interactions, in which other (heavy!) Higgs fields 
have nonzero vacuum expectation values that may give rise to still 
larger vacuum energies.  At a fundamental level, we can therefore conclude 
that a spontaneously broken gauge theory of the strong, weak, and 
electromagnetic interactions---or merely of the electroweak 
interactions---cannot be complete. The vacuum energy problem must be an important 
clue.  But to what?

The tentative evidence for  a nonzero cosmological constant recasts the problem in two important ways.
First, instead of looking for a principle that would forbid a cosmological constant, perhaps a symmetry principle that would set it exactly to zero, we may be called upon to explain a tiny cosmological constant.
Second, if the interpretation of the accelerating expansion in terms of dark energy is correct, we now have observational access to some new stuff whose equation of state and other properties we can try to measure.  Maybe that will give us the clues that we need to solve this old problem, and to understand how it relates to the electroweak theory.
  
 \section{Big Questions}
Opening the Fermi scale to exploration means entering a new world. The quest for the origins of electroweak symmetry breaking---testing the Higgs mechanism---heads what promises to be a very rich experimental agenda. What we learn from the new round of experimentation, and the interplay with theory, will transfigure particle physics and deepen our understanding of the everyday world. Moreover, we have the strong suspicion that many of the outstanding problems of particle physics and cosmology may be linked---and linked to the Fermi scale.

Here are some of the questions that will shape the explorations to come:
\begin{enumerate}
\item What is the agent that hides the electroweak symmetry?
\item Is there a Higgs boson? Might there be several?
\item Does the Higgs boson give mass to fermions, or only to the weak bosons? What sets the masses and mixings of the quarks and leptons?
\item How does the Higgs boson interact with itself? What shapes the Higgs potential?
\item Could we be living in a false (metastable) vacuum?
\item Is the Higgs boson elementary or composite?
\item Does the pattern of Higgs-boson decays imply new physics? Will uexpected or rare decays of the Higgs boson reveal new kinds of matter?
\item What stabilizes the Higgs-boson mass on the Fermi scale? Is Nature supersymmetric? Is electroweak symmetry breaking an emergent phenomenon connected with strong dynamics? Is electroweak symmetry breaking related to gravity through extra spacetime dimensions?
\item How can a light Higgs boson coexist with the absence of signals for new phenomena?
\item What resolves the vacuum energy problem?
\item What lessons does electroweak symmetry breaking hold for unified theories of the strong, weak, and electromagnetic interactions? for the inflationary Universe? for dark energy?
\end{enumerate}
It is an inspiring list. Within a decade, we can expect to have many answers---and even better questions!

\ack
Fermilab is operated by Fermi Research Alliance, LLC  under Contract No.~DE-AC02-07CH11359 with the United States Department of Energy.  It is a pleasure to thank Luis \'{A}lvarez-Gaum\'{e} and other members of the CERN Theory Group for warm hospitality in Geneva. I thank Abdelhak Djouadi and Olga Mena for providing figures, and Ian Aitchison, Michael Chanowitz, Mu-Chun Chen, JoAnne Hewett, J. D. Jackson, Andreas Kronfeld, Joe Lykken, Tom Rizzo, and Ruth Van de Water for helpful advice.

\bibliography{IoPHiggs}

\end{document}